\documentclass[aps,prb,reprint,superscriptaddress]{revtex4-2}

\usepackage{graphicx}
\usepackage{physics}
\usepackage{makecell}
\begin{document}

% Use the \preprint command to place your local institutional report
% number in the upper righthand corner of the title page in preprint mode.
% Multiple \preprint commands are allowed.
% Use the 'preprintnumbers' class option to override journal defaults
% to display numbers if necessary
%\preprint{}

%Title of paper
\title{Terahertz Faraday and Kerr rotation spectroscopy of $\text{Bi}_{1-x}\text{Sb}_{x}$ films \\in high magnetic fields up to 30~Tesla}

\normalsize

\author{Xinwei~Li}
\thanks{These authors contributed equally.}
\affiliation{Department of Electrical and Computer Engineering, Rice University, Houston, Texas 77005, USA}

\author{Katsumasa~Yoshioka}
\thanks{These authors contributed equally.}
\affiliation{Department of Physics, Graduate School of Engineering, Yokohama National University, Yokohama 240-8501, Japan}

\author{Ming~Xie}
\thanks{These authors contributed equally.}
\affiliation{Department of Physics and Center for Complex Quantum Systems, The University of Texas at Austin, Austin, Texas 78712, USA}

\author{G.~Timothy~Noe~II}
\affiliation{Department of Electrical and Computer Engineering, Rice University, Houston, Texas 77005, USA}

\author{Woojoo Lee}
\affiliation{Department of Physics and Center for Complex Quantum Systems, The University of Texas at Austin, Austin, Texas 78712, USA}

\author{Nicolas~Marquez~Peraca}
\affiliation{Department of Physics and Astronomy, Rice University, Houston, Texas 77005, USA}

\author{Weilu~Gao}
\affiliation{Department of Electrical and Computer Engineering, Rice University, Houston, Texas 77005, USA}

\author{Toshio~Hagiwara}
\affiliation{Department of Physics, Graduate School of Engineering, Yokohama National University, Yokohama 240-8501, Japan}

\author{Handegard~S.~Orjan}
\affiliation{National Institute for Materials Science, Tsukuba, Ibaraki 305-0044, Japan}

\author{Li-Wei~Nien}
\affiliation{National Institute for Materials Science, Tsukuba, Ibaraki 305-0044, Japan}

\author{Tadaaki~Nagao}
\affiliation{National Institute for Materials Science, Tsukuba, Ibaraki 305-0044, Japan}

\author{Masahiro~Kitajima}
\affiliation{Department of Physics, Graduate School of Engineering, Yokohama National University, Yokohama 240-8501, Japan}
\affiliation{National Institute for Materials Science, Tsukuba, Ibaraki 305-0044, Japan}

\author{Hiroyuki~Nojiri}
\affiliation{Institute for Materials Research, Tohoku University, Sendai 980-8577, Japan}

%\author{Xiaoqin~Li}
%\affiliation{Department of Physics and Center for Complex Quantum Systems, The University of Texas at Austin, Austin, Texas 78712, USA}

\author{Chih-Kang~Shih}
\affiliation{Department of Physics and Center for Complex Quantum Systems, The University of Texas at Austin, Austin, Texas 78712, USA}

\author{Allan~H.~MacDonald}
\affiliation{Department of Physics and Center for Complex Quantum Systems, The University of Texas at Austin, Austin, Texas 78712, USA}

\author{Ikufumi~Katayama}
%\email{katayama-ikufumi-bm@ynu.ac.jp}
\affiliation{Department of Physics, Graduate School of Engineering, Yokohama National University, Yokohama 240-8501, Japan}

\author{Jun~Takeda}
%\email{jun@ynu.ac.jp}
\affiliation{Department of Physics, Graduate School of Engineering, Yokohama National University, Yokohama 240-8501, Japan}

\author{Gregory~A.~Fiete}
%\email{g.fiete@northeastern.edu}
\affiliation{Department of Physics and Center for Complex Quantum Systems, The University of Texas at Austin, Austin, Texas 78712, USA}
\affiliation{Department of Physics, Northeastern University, Boston, Massachusetts 02115, USA}
\affiliation{Department of Physics, Massachusetts Institute of Technology, Cambridge, Massachusetts 02139, USA}

\author{Junichiro~Kono}
%\email{kono@rice.edu}
\affiliation{Department of Electrical and Computer Engineering, Rice University, Houston, Texas 77005, USA}
\affiliation{Department of Physics and Astronomy, Rice University, Houston, Texas 77005, USA}
\affiliation{Department of Material Science and NanoEngineering, Rice University, Houston, Texas 77005, USA}

\date{\today}

\begin{abstract}
%Surface carriers in a three-dimensional topological insulator, which are characterized by gapless spin-momentum-locked surface bands within a bulk band gap, are promising for robust and low-dissipation electronics and spintronics. However, as most topological insulators suffer from unintentional doping, separating the electronic contributions from the surface and the bulk remains challenging. 

%Magneto-optical spectroscopy has emerged as an important tool to study carrier dynamics in three-dimensional topological insulators. While a high magnetic field is advantageous to bring the system to the quantum limit, previous studies have all been working in the intermediate magnetic field range. Here, 
We report results of terahertz Faraday and Kerr rotation spectroscopy measurements on thin films of $\text{Bi}_{1-x}\text{Sb}_{x}$, an alloy system that exhibits a semimetal-to-topological-insulator transition as the Sb composition $x$ increases.  By using a single-shot time-domain terahertz spectroscopy setup combined with a table-top pulsed mini-coil magnet, we conducted measurements in magnetic fields up to 30~T, observing distinctly different behaviors between semimetallic ($x < 0.07$) and topological insulator ($x > 0.07$) samples.  Faraday and Kerr rotation spectra for the semimetallic films showed a pronounced dip that blue-shifted with the magnetic field, whereas spectra for the topological insulator films were positive and featureless, increasing in amplitude with increasing magnetic field and eventually saturating at high fields ($>$20~T). Ellipticity spectra for the semimetallic films showed resonances, whereas the topological insulator films showed no detectable ellipticity.  To explain these observations, we developed a theoretical model based on realistic band parameters and the Kubo formula for calculating the optical conductivity of Landau-quantized charge carriers.  Our calculations quantitatively reproduced all experimental features, establishing that the Faraday and Kerr signals in the semimetallic films predominantly arise from bulk hole cyclotron resonances while the signals in the topological insulator films represent combined effects of surface carriers originating from multiple electron and hole pockets.  These results demonstrate that the use of high magnetic fields in terahertz magnetopolarimetry, combined with detailed electronic structure and conductivity calculations, allows us to unambiguously identify and quantitatively determine unique contributions from different species of carriers of topological and nontopological nature in Bi$_{1-x}$Sb$_x$.

%The Faraday rotation induced by the semimetal films had a pronounced negative peak which blueshifted with the magnetic field, while that induced by the topological insulator film appeared to be positive, was spectrally featureless, and first increased then saturated with the magnetic field. 
\end{abstract}

% insert suggested keywords - APS authors don't need to do this
%\keywords{}

%\maketitle must follow title, authors, abstract, and keywords

\maketitle

\section{Introduction}

One of the most appealing methods proposed for probing topologically protected surface states in a three-dimensional (3D) topological insulator (TI) is observing the topological magnetoelectric (TME) effect~\cite{QietAl08PRB,EssinetAl09PRL,TseMacDonald10PRL,MaciejkoetAl10PRL,TseMacDonald11PRB,WangetAl15PRB,MorimotoetAl15PRB,ZhangetAl19PRL}. When time-reversal symmetry is broken by a magnetic field, a gap appears in the surface Dirac state(s), and the Hall conductivity of the sample surface is half-integer quantized, which in turn leads to a quantized bulk magnetoelectric response. This phenomenon can be described through an electromagnetic Lagrangian analogous to the theory of axion electrodynamics in particle physics~\cite{Wilczek87PRL}. %Although various proposals exist to experimentally detect the topological magnetoelectric effect, 
Experimentally, the most common and established experimental techniques for studying 3D TIs are angle-resolved photoemission spectroscopy (ARPES)~\cite{HsiehetAl08Nature,HsiehetAl09Science,XiaetAl09NP,HsiehetAl09Nature,HiraharaetAl10PRB,NishideetAl10PRB,NakamuraetAl11PRB,BeniaetAl15PRB,LeeetAl17EPL} and electronic transport~\cite{LenoiretAl96JPCS,TaskinAndo09PRB,TaskinetAl10PRB,AnalytisetAl10PRB,ButchetAl10PRB,RenetAl10PRB,QuetAl10Science} measurements, both of which are not suited for detecting the TME effect. Specifically, a magnetic field cannot be applied in ARPES experiments, and there are limitations due to nontopological edge states in transport experiments.

Magneto-optical spectroscopy at low photon energies, such as the microwave and far-infrared spectral ranges, has been recognized to be an ideal probe for the TME effect~\cite{MaciejkoetAl10PRL,TseMacDonald10PRL,TseMacDonald11PRB}. By using electromagnetic radiation whose photon energy is comparable to or smaller than the magnetic-field-induced surface gap, previous terahertz (THz) magneto-optical spectroscopy experiments have detected changes of the surface optical conductivity induced by an applied small or moderate magnetic field~\cite{JenkinsetAl10PRB,LaForgeetAl10PRB,HancocketAl11PRL,ValdesAguilaretAl12PRL,WuetAl15PRL,WuetAl16Science,DziometAl17NC}.  Applying a much stronger magnetic field on the sample is, generally speaking, expected to be more advantageous because the field can induce a gap much larger than the photon energy and bring the surface carrier states toward the quantum limit~\cite{ZhuetAl17NC}. %However, low-energy optical spectroscopy has never been equipped with the capability to apply a high magnetic field to study a 3D TI.

\begin{figure}[b]
	\begin{center}
		\includegraphics[scale=0.47]{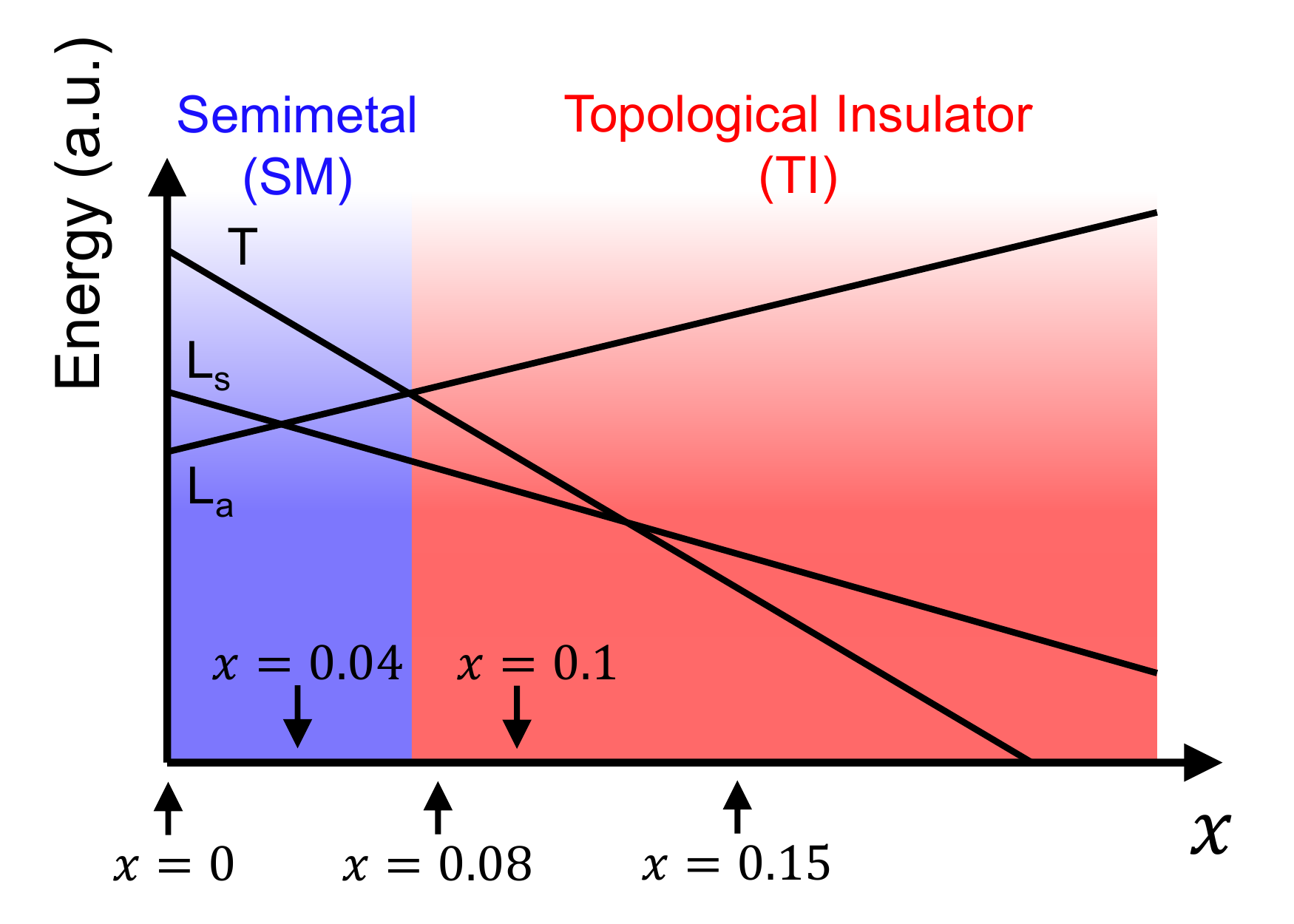}
		\caption{\small Schematic phase diagram of $\text{Bi}_{1-x}\text{Sb}_{x}$ alloys.  The energies of different band edges at high symmetry points (T and L points) are plotted as a function of Sb composition $x$.  The five arrows pointing to the horizontal axis mark the nominal values of $x$ of the five film samples we studied.  See Table \ref{BiSbtable} for more details of the characteristics of the samples.}
		\label{Phasediagram}
	\end{center}
\end{figure}

In this article, we describe results of THz Faraday and Kerr rotation spectroscopy measurements on thin films of $\text{Bi}_{1-x}\text{Sb}_{x}$. This alloy system is known to show a semimetal (SM)-to-TI transition as a function of $x$, and its phase diagram (Fig.\,\ref{Phasediagram}) is well established~\cite{FuKane07PRB,TeoetAl08PRB,ZhangetAl09PRB,HsiehetAl08Nature,HsiehetAl09Nature}, although some controversy remains as to whether surface states in the SM region are topologically nontrivial~\cite{ItoetAl16PRL}.  We performed measurements using a single-shot THz time-domain polarimetry setup combined with a 30-T pulsed magnet system~\cite{NoeetAl13RSI,NoeetAl14AO,NoeetAl16OE}.  The wide tunability of the band structure of $\text{Bi}_{1-x}\text{Sb}_{x}$ with $x$ makes this material system suitable for identifying and distinguishing the uniquely different magneto-optical responses of samples in the topologically trivial and nontrivial phases.

A THz beam normally incident on the sample surface in the Faraday geometry exhibited polarization rotations due to the field-induced Hall conductivities of the surface and/or bulk carriers. Faraday rotation spectra for the SM films had a pronounced dip, which blue-shifted with the magnetic field, while the Faraday rotations in the TI films were positive and spectrally featureless, increasing and then saturating with increasing magnetic field.  Using a theoretical model incorporating realistic band parameters and the Kubo formula for calculating the optical conductivity, we found that the optical Hall signals in the SM (TI) samples can be attributed to carriers in the bulk (surface) bands.  The model suggested that the magneto-optical signal from the SM films was dominated by the cyclotron resonance of bulk high-mobility holes, while that from the TI film resulted from the summed contributions of multiple electron and hole pockets associated with the surface bands. %Our work introduces the technique that combines THz polarimetry with pulsed high magnetic fields to the study of 3D TIs, and provides a way to distinguish the origin of optical conductivity signals from topological materials using the theoretical modeling.

%%%%%%%%%%%%%%%%%%%%%%%%%%%%%%%%%%%
%%%%%%%%%%%%%%%%%%%%%%%%%%%%%%%%%%%
\section{Samples and Methods}
%%%%%%%%%%%%%%%%%%%%%%%%%%%%%%%%%%%
%%%%%%%%%%%%%%%%%%%%%%%%%%%%%%%%%%%

\subsection{$\text{Bi}_{1-x}\text{Sb}_{x}$ films}

We studied $\text{Bi}_{1-x}\text{Sb}_{x}$ films on silicon substrates grown by molecular beam epitaxy using the methods described in Refs.\,\onlinecite{HiraharaetAl10PRB,KatayamaetAl18PRB}. Five samples with different $x$ values were studied; see Table \ref{BiSbtable}. The nominal $x$ values were 0, 0.04, 0.08, 0.1, and 0.15, respectively, while the thickness, $t$, was nominally 40~nm for all films. According to Fig.\,\ref{Phasediagram}, Samples 1 and 2 were in the SM regime while Samples 3--5 were in the TI regime.  We used a combination of structural and chemical characterization methods to precisely determine the actual values of $x$, $t$, and crystal orientation of the films; \textit{in-situ} reflection high-energy electron diffraction (RHEED) patterns determined the crystal orientation, while \textit{ex-situ} x-ray diffraction (XRD), x-ray fluorescence (XRF), and atomic force microscopy (AFM) experiments provided information on the crystal structure, chemical composition, and film thickness, respectively.  The obtained parameters of the samples are summarized in Table.\,\ref{BiSbtable}.

\begin{table*}
\noindent\centering\begin{tabular}{c||c|c|c|c|c}
	Sample No. & Nominal $x$ & Actual $x$ & Thickness $t$ (nm) & Orientation & Character\\\hline \hline
	1 & 0 & 0 & 68 (AFM) & $\left < 001 \right >$ (RHEED) & SM\\\hline
	2 & 0.04 & 0.03 (XRF) & 60 (XRF) & $\left < 001 \right >$ (XRD) & SM\\\hline
	3 & 0.08 & $0.03<x<0.136$  & 70 (AFM) & N/A & TI\\\hline
	4 & 0.10 & 0.136 (XRF) & 54 (XRF) & \makecell{$\left < 001 \right >$ and $\left < 012 \right >$ (XRD) \\ $\left < 001 \right >$ (ARPES)} & TI \\\hline
	5 & 0.15 & $x>0.136$ & 77 (AFM) & $\left < 001 \right >$ (RHEED) & TI\\\hline
\end{tabular}
\caption{Characteristics of the five $\text{Bi}_{1-x}\text{Sb}_{x}$ thin film samples studied.  The Sb content ($x$), film thickness ($t$), and crystal orientation of the films are shown.  The experimental methods used to determine the parameters are indicated in the parentheses. SM = semimetal.  TI = topological insulator.}
\label{BiSbtable}
\end{table*}

For Sample 1 ($x=0$), RHEED determined that the film orientation was $\left < 001 \right >$. No Sb was incorporated in this sample, so that chemical analysis was not needed. AFM determined that $t$ = 68~nm. 
For Sample 2 (nominal $x$ = 0.04), XRF and XRD measurements were performed.  An obtained XRF spectrum was fit with a model built in the measurement software, using $x$ and $t$ as adjustable parameters, and the parameters that gave the best fit were $x=0.03$ and $t$ = 60~nm. XRD showed a dominating diffraction peak due to the (001) plane of the film, confirming that the crystal orientation was $\left < 001 \right >$.

\begin{figure}[htb]
	\begin{center}
		\includegraphics[width=0.95\linewidth]{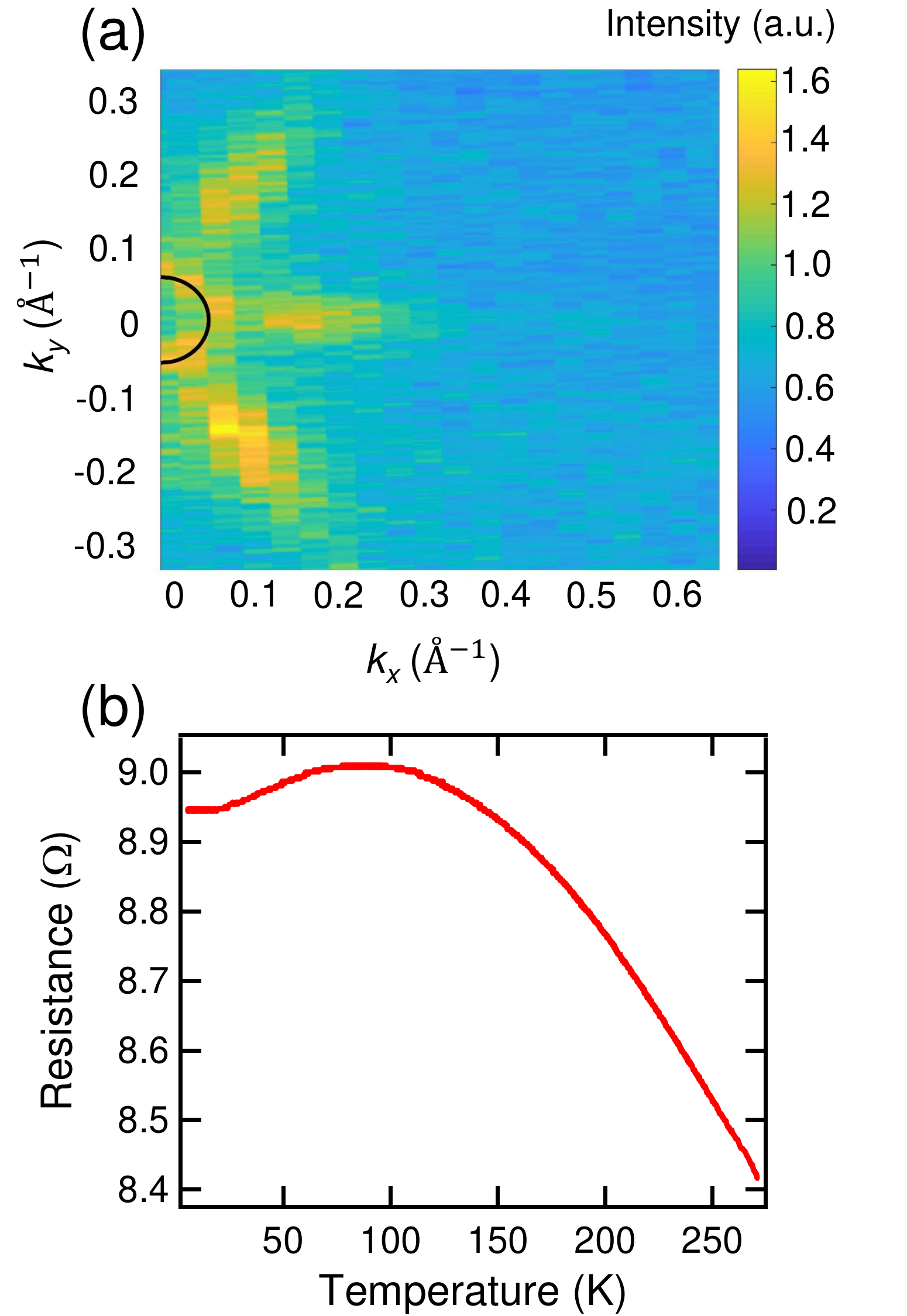}
		\caption{\small Electronic properties of Sample 4. (a)~Fermi surface mapping using ARPES. Black solid line marks the fitting of Fermi energy by assuming a Dirac velocity of $2600\ \text{meV\AA}$. (b)~Resistance versus temperature characteristic. }
		\label{Bi090electronic}
	\end{center}
\end{figure}

For Sample 4 (nominal $x$ = 0.10), fitting analysis on an XRF spectrum allowed us to determine $x=0.136$ and $t$ = 54~nm. An XRD curve showed diffraction peaks from the (001) and (012) planes, suggesting that the film possibly had some spatial inhomogeneity in terms of orientation; however, the ARPES data shown in Fig.\,\ref{Bi090electronic}(a) confirmed that the film area on which our magneto-optical measurements were performed was dominated by the $\left < 001 \right >$ orientation. Transport measurements were also performed on the film; see Fig.\,\ref{Bi090electronic}(b) for the resistance-temperature ($R$-$T$) characteristic. The increasing $R$ with decreasing $T$ in the $120$~K $<T<$ $250$~K region can be attributed to the decreasing number of thermal carriers in the insulating bulk, while the subsequent decrease of $R$ in the $T<120$~K region is likely due to surface metallicity; such behavior has been observed in $R$-$T$ curves measured for TI systems with minimal bulk doping~\cite{RenetAl10PRB,QuetAl10Science}.

For Samples 3 (nominal $x$ = 0.08) and 5 (nominal $x$ = 0.15), we found that the uncertainties given by model fits to their XRF spectra were too large to determine both $x$ and $t$ accurately.  %Therefore, we could not confirm that these two films were actually in the TI phase. 
However, as the amount of Sb incorporation was progressively increased in growing the five films, from Sample 1 to Sample 5, it is certain that Sample 3 had an $x$ value that is between those of Samples 2 and 4 while Sample 5 had $x>0.136$.

\subsection{THz polarimetry}\label{polarimetry}

\begin{figure}[htb]
	\begin{center}
		\includegraphics[scale=0.25]{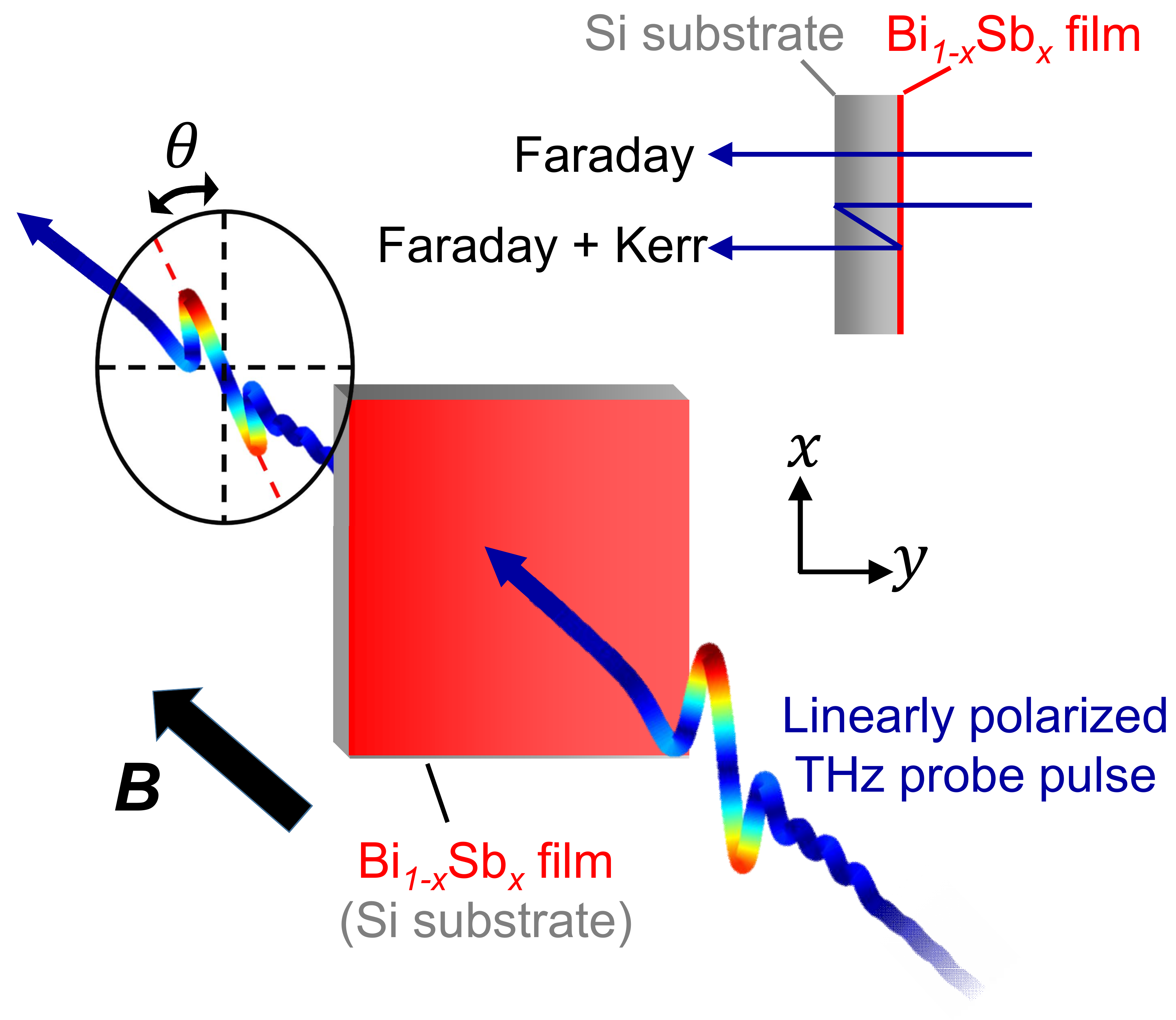}
		\caption{\small THz polarimetry setup probing the magneto-optical response of the $\text{Bi}_{1-x}\text{Sb}_{x}$ films. Both Faraday and Kerr rotation angles can be measured with the transmission geometry. }
		\label{BiSbsetup}
	\end{center}
\end{figure}

Figure~\ref{BiSbsetup} schematically shows the THz polarimetry technique we used to probe the magneto-optical response of the $\text{Bi}_{1-x}\text{Sb}_{x}$ films; similar techniques have previously been used to study systems other than TIs~\cite{IkebeetAl10PRL,ShimanoetAl13NC,OkadaetAl16NC}. Our laser system was a Ti:sapphire regenerative amplifier (1~kHz, 150~fs, 775~nm, Clark-MXR, inc.), which generated and detected THz probe pulses with ZnTe crystals. The incident THz beam was linearly polarized in the $x$ direction, but both the $x$ and $y$ components of the electric field ($E_x$ and $E_y$) of the transmitted field were measured by electro-optic sampling in the time domain to quantify the THz polarization state that was affected by the carrier Hall effect. Because the THz waveforms were both amplitude- and phase-resolved, we were able to detect polarization rotations as small as 1~mrad as well as the corresponding ellipticity change. In addition, the pulse-based technique allowed us to measure both Faraday and Kerr rotations using only the transmission geometry. 

As shown in the upper right schematic in Fig.\,\ref{BiSbsetup}, the THz pulse that directly passes through the film and the substrate gives the Faraday rotation. However, a portion of the pulse experiences additional reflection events at the vacuum-substrate interface and the film-vacuum interface, and can still be measured as a back reflection pulse in the same alignment geometry; the back reflection pulse appears later in time than the main pulse that is directly transmitted due to its additional path inside the substrate, but it contains both Faraday and Kerr rotation signals. The Kerr rotation signal can be isolated out by subtracting the Faraday rotation from the total polarization rotation of the back reflection pulse.

Below, we show the process of determining the polarization state of a pulse. We start with measurements of the $E_x(t)$ and $E_y(t)$ of a pulse in the time domain. Since the $E_y(t)$ signal is usually much smaller than $E_x(t)$, we changed the polarity of the magnetic field $B$ and took the average of the subtraction $[E_y(+B)-E_y(-B)]/2$ as the real $E_y(t)$ signal that excludes any effect of imperfect linear polarization of the incident beam. We Fourier-transformed $E_x(t)$ and $E_y(t)$ into complex-valued $E_x(\omega)$ and $E_y(\omega)$, which are functions of frequency $\omega$. We then defined the modes in the circular polarization basis,
\begin{equation}
E_{\pm}(\omega)=\frac{1}{\sqrt{2}}[E_x(\omega)\pm iE_y(\omega)].
\end{equation}
Then, the rotation of the polarization plane and change of ellipticity can be quantified, respectively, as
\begin{align}
\theta(\omega)= & \frac{\text{arg}[E_+(\omega)]-\text{arg}[E_-(\omega)]}{2},\\
\eta(\omega)= & \frac{|E_-(\omega)|-|E_+(\omega)|}{|E_-(\omega)|+|E_+(\omega)|}.
\label{eq:rotell}
\end{align}
In addition, the real and imaginary parts of the complex-valued longitudinal and Hall conductivity of the film can be determined if a reference pulse signal $E_\text{ref}$ is collected as a THz pulse passed through the bare substrate without the film. We first calculated the complex-valued conductivity in the circular basis
\begin{equation}
\sigma_\pm(\omega)=\frac{(1+n_\text{Si})(1-t_\pm)}{Z_0dt_\pm},
\end{equation}
where $n_\text{Si}=3.5$ is the refractive index of the silicon substrate, $Z_0=377\ \Omega$ is the vacuum impedance, $d$ is the thickness of the film, and $t_\pm=E_\pm(\omega)/E_{\text{ref,}\pm}$. Then, the longitudinal and Hall conductivities of the film can be obtained, respectively, as
\begin{align}
\sigma_{xx}= & \frac{\sigma_+(\omega)+\sigma_-(\omega)}{2},\\
\sigma_{xy}= & \frac{-\sigma_+(\omega)+\sigma_-(\omega)}{2i}.
\label{sigma_xy}
\end{align}
Just like the complex-valued longitudinal and Hall conductivity ($\sigma_{xx}$ and $\sigma_{xy}$), the combination of the polarization rotation angle and ellipticity change ($\theta$ and $\eta$) contains complete information on the THz magneto-optical response of the film, with the only difference that a reference signal is not needed in determining $\theta$ and $\eta$. 

\begin{figure}[htb]
	\begin{center}
		\includegraphics[scale=0.55]{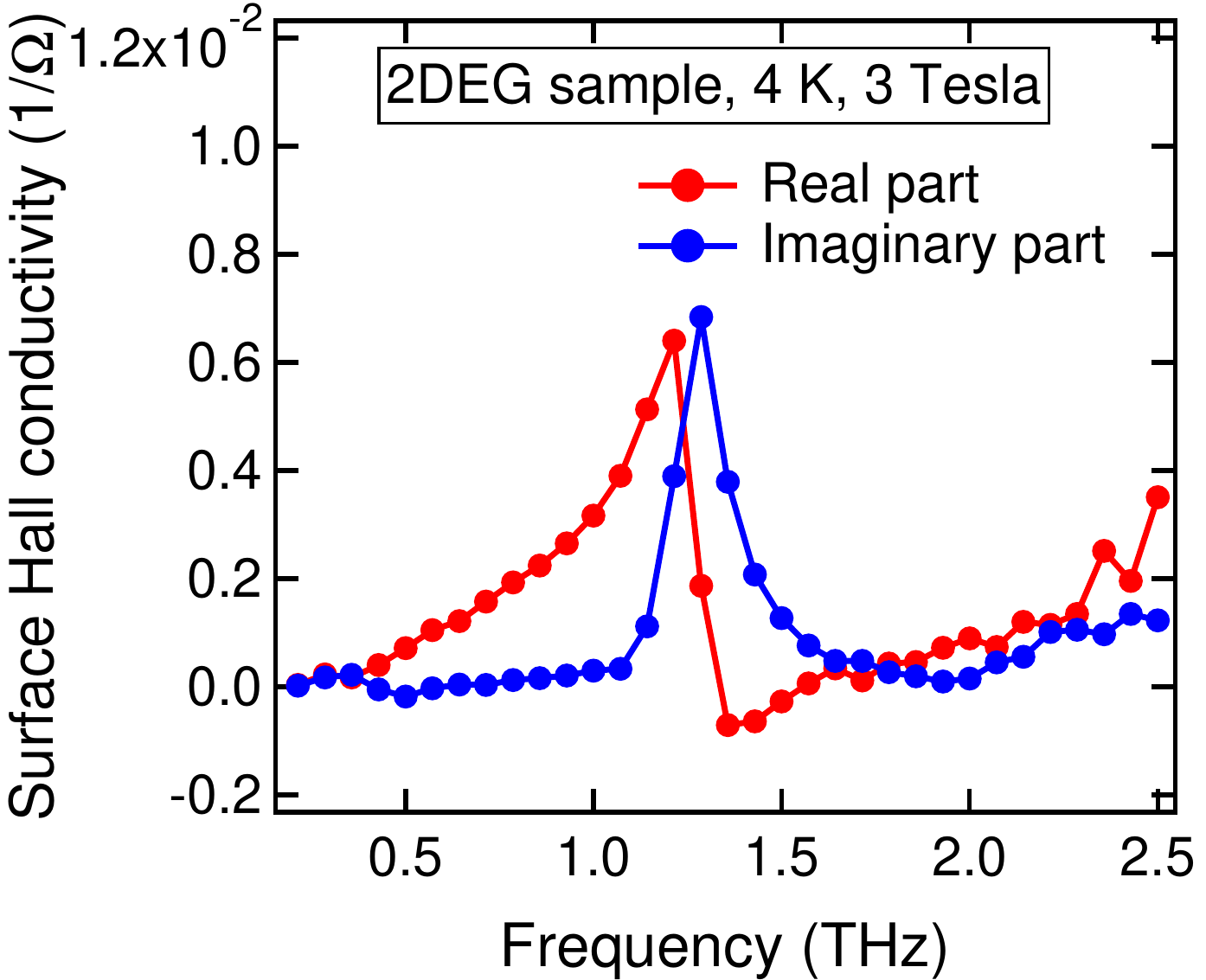}
		\caption{\small Surface Hall conductivity of a standard GaAs 2DEG sample at 3~T. The electron cyclotron resonance appears as a positive peak, which was used to determine the dominant carrier type in the $\text{Bi}_{1-x}\text{Sb}_{x}$ samples.}
		\label{2DEGHall}
	\end{center}
\end{figure}

Electronic properties of a sample, irrespective of whether it is a SM or TI, contain a mixture of contributions from electrons and holes. Sometimes one carrier type can be dominant, and determining the carrier type gives valuable information on the origin of the optical conductivity signal. We can use the sign of the Hall conductivity to determine whether the dominant carriers are electrons or holes. We performed THz transmission measurements on a standard GaAs two-dimensional electron gas (2DEG) sample at 3~T, where the electron cyclotron resonance peak appeared in the middle of the THz spectrum, and calculated its optical Hall conductivity. As shown in Fig.\,\ref{2DEGHall}, both the real and imaginary parts of the Hall conductivity are positive at the cyclotron resonance peak. We used this sign as a reference to determine the dominant carrier type in the $\text{Bi}_{1-x}\text{Sb}_{x}$ samples.
  
\subsection{Single-shot THz spectroscopy in pulsed high magnetic fields}\label{RAMBO}

\begin{figure}[htb]
	\begin{center}
		\includegraphics[scale=0.42]{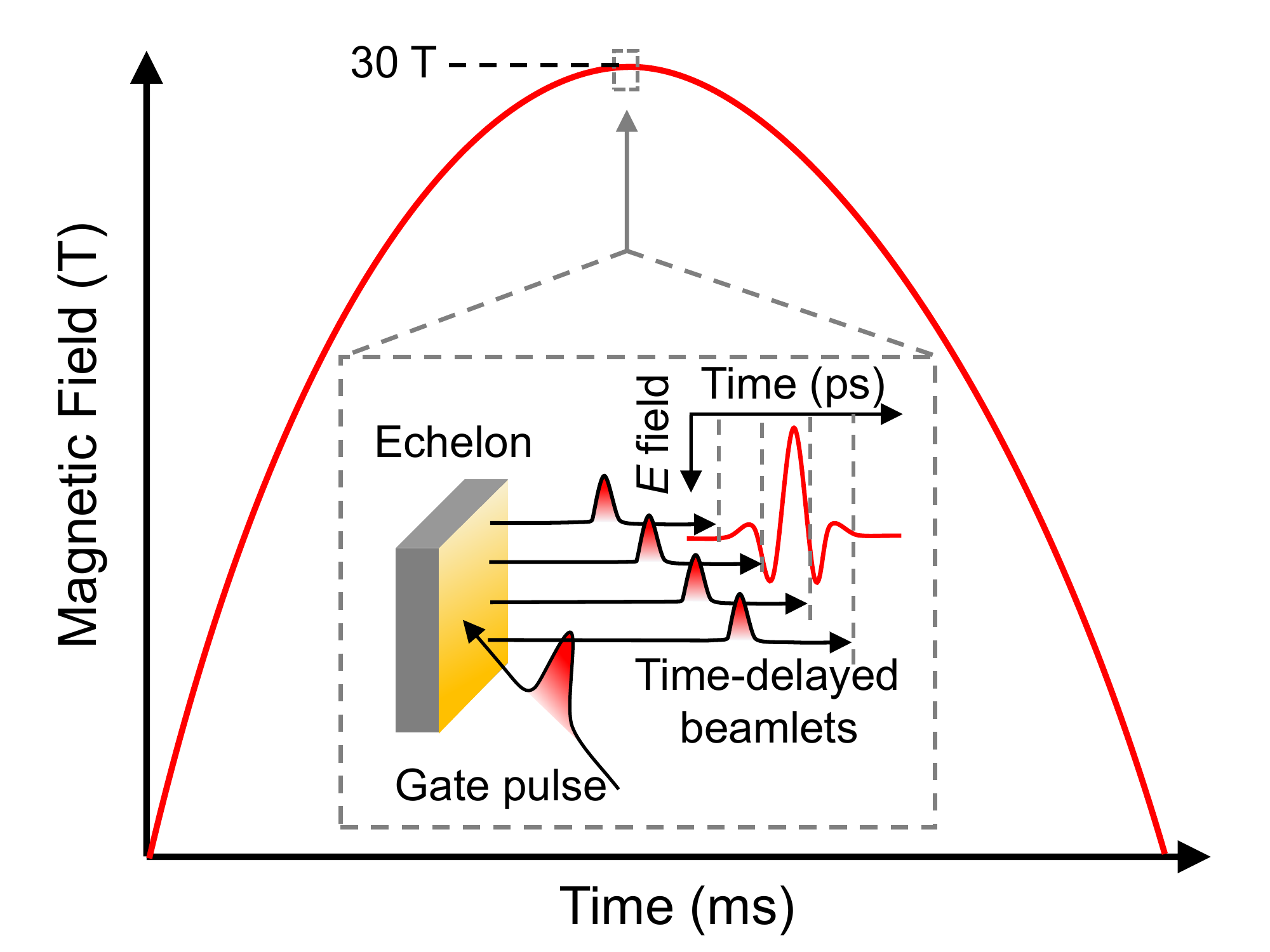}
		\caption{\small Schematic diagram of the single-shot THz spectroscopy system. At the peak of the pulsed magnetic field, an optical gate pulse incident onto a reflective echelon mirror is converted into time-delayed beamlets, which enable mapping out the time-domain THz electric field waveform in a single shot without involving any delay stage.}
		\label{RAMBOTHz}
	\end{center}
\end{figure}

In addition to using a standard delay-stage-based step-scan THz setup combined with a 10-T superconducting magnet~\cite{WangetAl07OL,WangetAl10NP,ArikawaetAl11PRB,ArikawaetAl12OE,ZhangetAl14PRL,ZhangetAl16NP,LietAl18NP,LietAl18Science}, we used a unique single-shot THz setup to perform measurements in pulsed magnetic fields up to 30~T~\cite{NoeetAl16OE}. Pulsed magnets typically generate higher magnetic fields with a smaller magnet and cryostat size than DC magnets, but the challenge is that the peak magnetic field only lasts for a short time (approximately 400~$\mu$s in our case~\cite{NoeetAl13RSI}), so that the time for a delay stage to step through the THz waveform is insufficient. We developed a single-shot THz time-domain detector, which is capable of measuring the full THz time-domain waveform using just one laser pulse within the 400-$\mu$s-long time window during which the sample is experiencing the peak of the pulsed magnetic field.

Single-shot THz detection relied on a reflective echelon mirror tilting the pulse front of the optical gate beam by forming time-delayed beamlets, which encoded time delay information across the beam intensity profile~\cite{KatayamaetAl11JJAP,MinamietAl13APL,MinamietAl15APL,MeadetAl19RSI}; see the inset in Fig.\,\ref{RAMBOTHz}. The temporal and spatial overlap of the gate beam and the THz beam was achieved at the electro-optical crystal. A combination of a quarter-wave plate, a Wollaston prism, and a silicon complementary metal-oxide-semiconductor (CMOS) camera was used to detect the THz-induced change of the gate beam polarization. The CMOS camera captured the full image of the gate beam, which contained information about the THz electric field at various time delays. The difference between the two images spatially separated on the CMOS camera by the Wollaston prism gave the time-domain THz electric field signal.

%%%%%%%%%%%%%%%%%%%%%%%%%%%%%%%%%%%
%%%%%%%%%%%%%%%%%%%%%%%%%%%%%%%%%%%
\section{Results}
%%%%%%%%%%%%%%%%%%%%%%%%%%%%%%%%%%%
%%%%%%%%%%%%%%%%%%%%%%%%%%%%%%%%%%%

\subsection{Semimetallic samples: Samples 1 and 2}

\begin{figure}[htb]
	\begin{center}
		\includegraphics[scale=0.4]{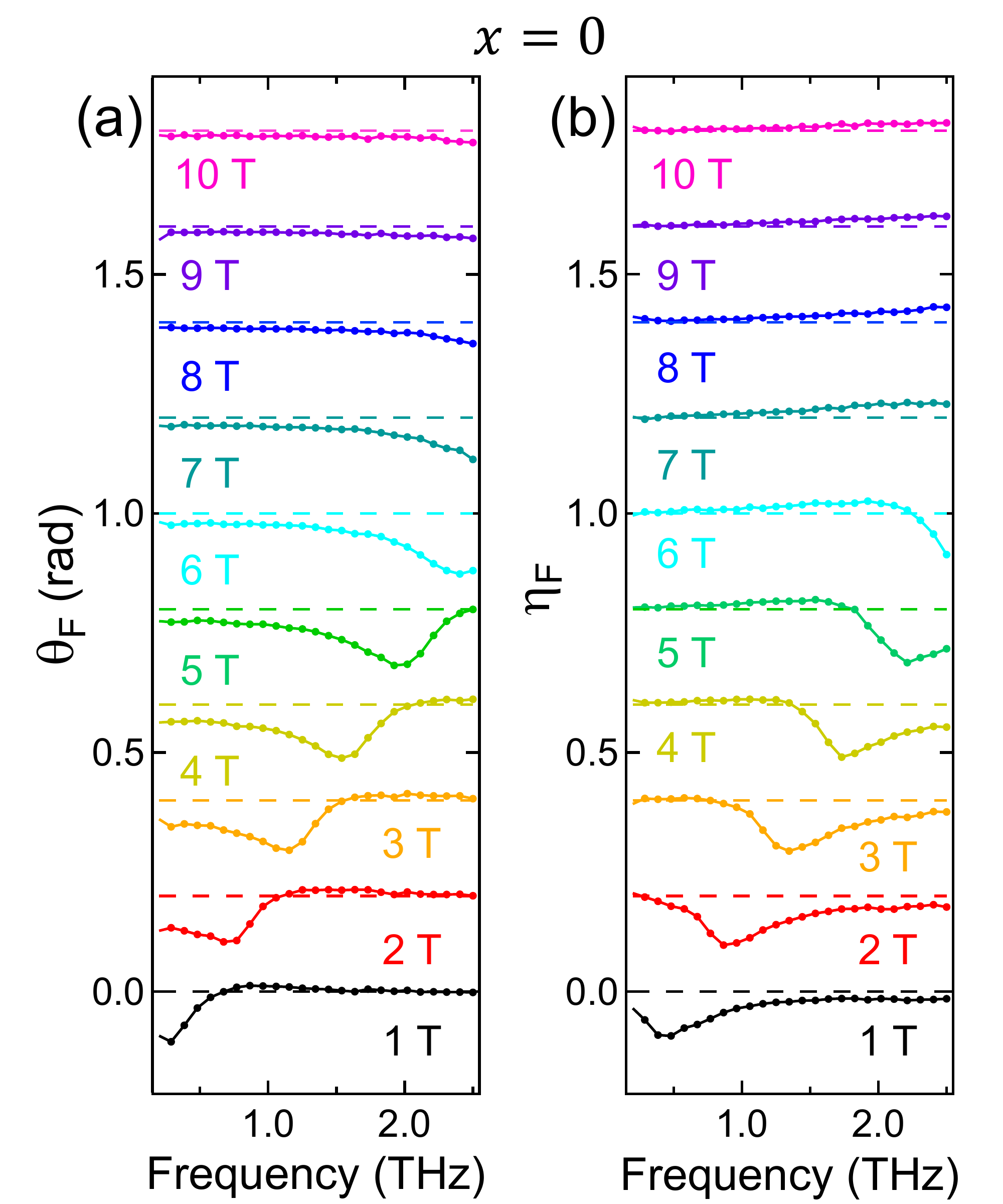}
		\caption{\small (a)~Faraday rotation and (b)~Faraday ellipticity spectra for Sample 1 at a temperature of 2~K at different magnetic fields up to 10~T.  Curves at different magnetic fields are vertically offset for clarity, and the baselines for the different spectra are indicated by dashed lines.}
		\label{Biexpt}
	\end{center}
\end{figure}

\begin{figure}[bt]
	\begin{center}
		\includegraphics[scale=0.4]{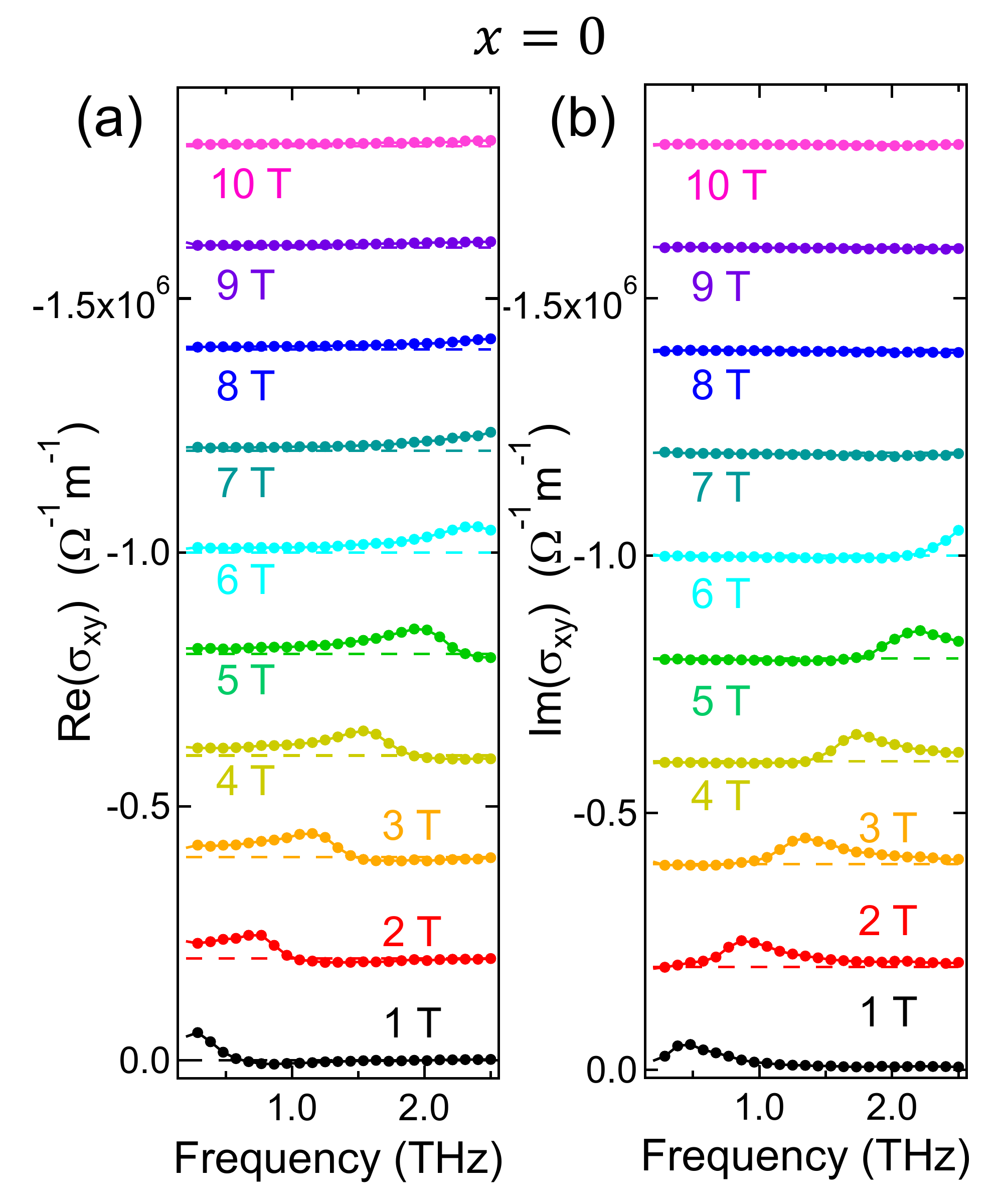}
		\caption{\small (a)~$\text{Re}(\sigma_{xy})$ and (b)~$\text{Im}(\sigma_{xy})$ spectra for Sample 1 at a temperature of 2~K at different magnetic fields up to 10~T.  Curves at different magnetic fields are vertically offset for clarity, and the baselines for the different spectra are indicated by dashed lines.}
		\label{Bisigma}
	\end{center}
\end{figure}

\begin{figure}[bt]
	\begin{center}
		\includegraphics[scale=0.4]{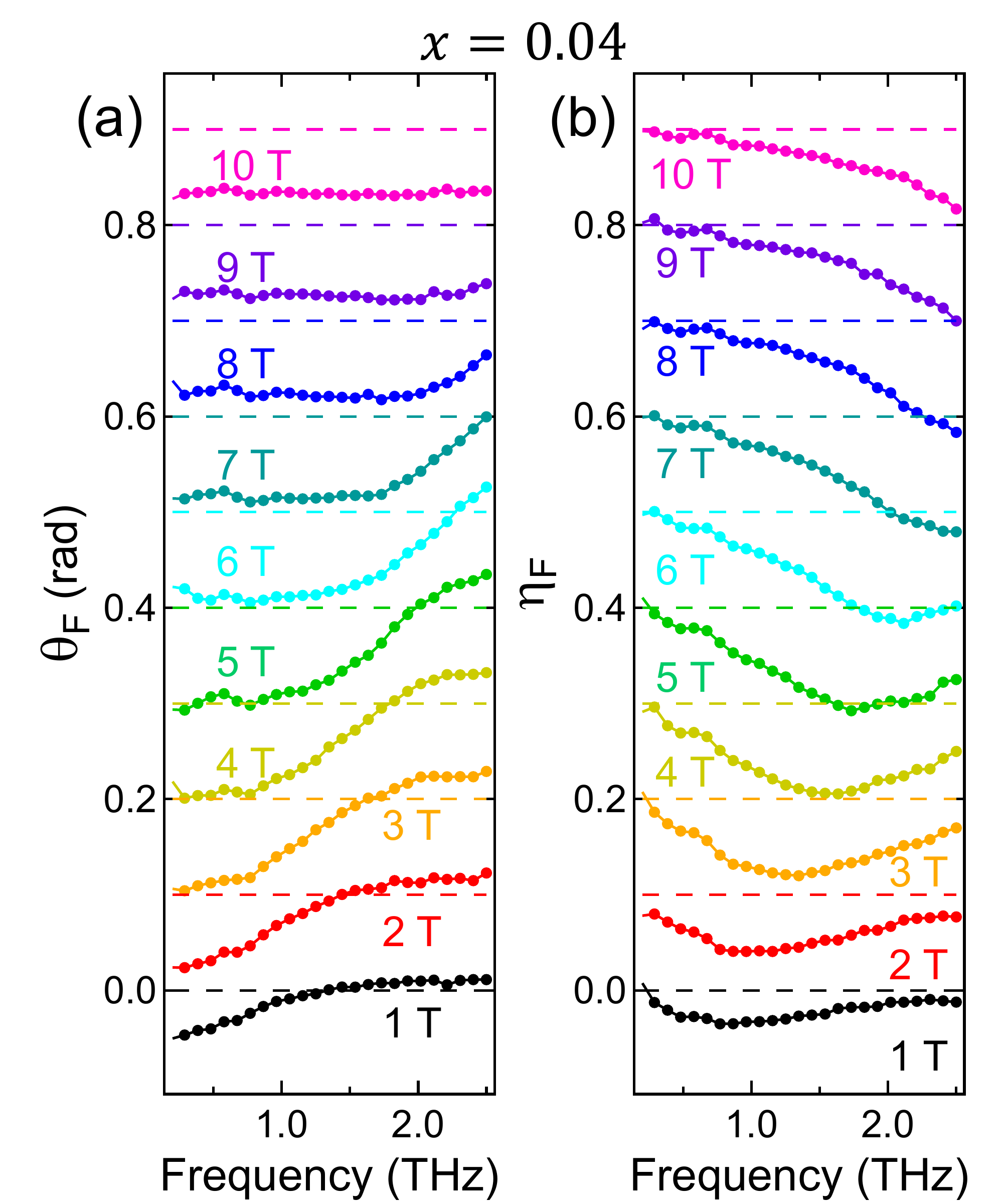}
		\caption{\small (a)~Faraday rotation and (b)~Faraday ellipticity spectra for Sample 2 at a temperature of 2~K at different magnetic fields up to 10~T.  Curves at different magnetic fields are vertically offset for clarity, and the baselines for the different spectra are indicated by dashed lines.}
		\label{Bi096Faraday}
	\end{center}
\end{figure}

\begin{figure}[htb]
	\begin{center}
		\includegraphics[scale=0.4]{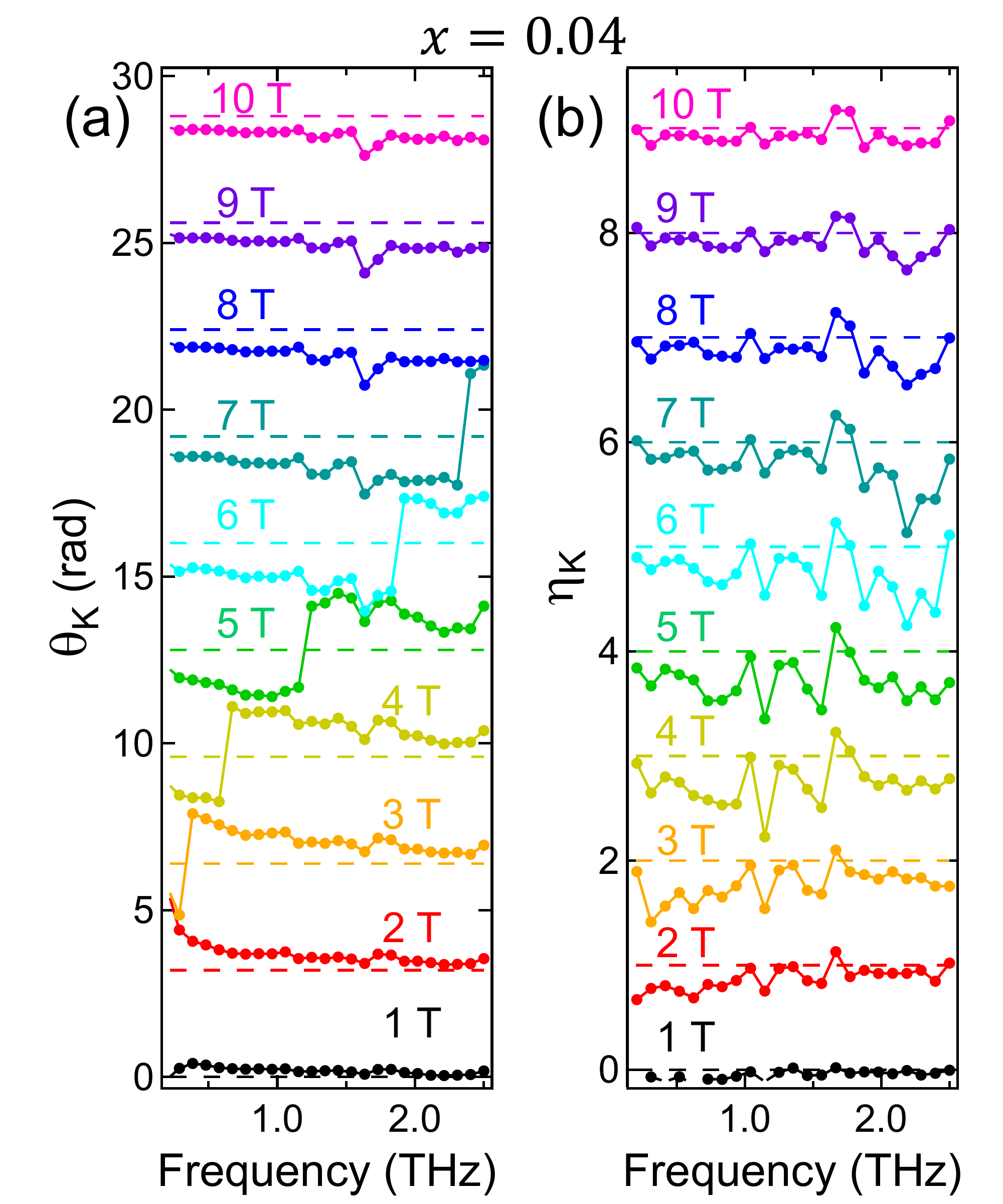}
		\caption{\small (a)~Kerr rotation and (b)~Kerr ellipticity spectra for Sample 2 at a temperature of 2~K at different magnetic fields up to 10~T. Curves at different magnetic fields are vertically offset for clarity, and the baselines for the different spectra are indicated by dashed lines.}
		\label{Bi096Kerr}
	\end{center}
\end{figure}

\begin{figure}[htb]
	\begin{center}
		\includegraphics[scale=0.65]{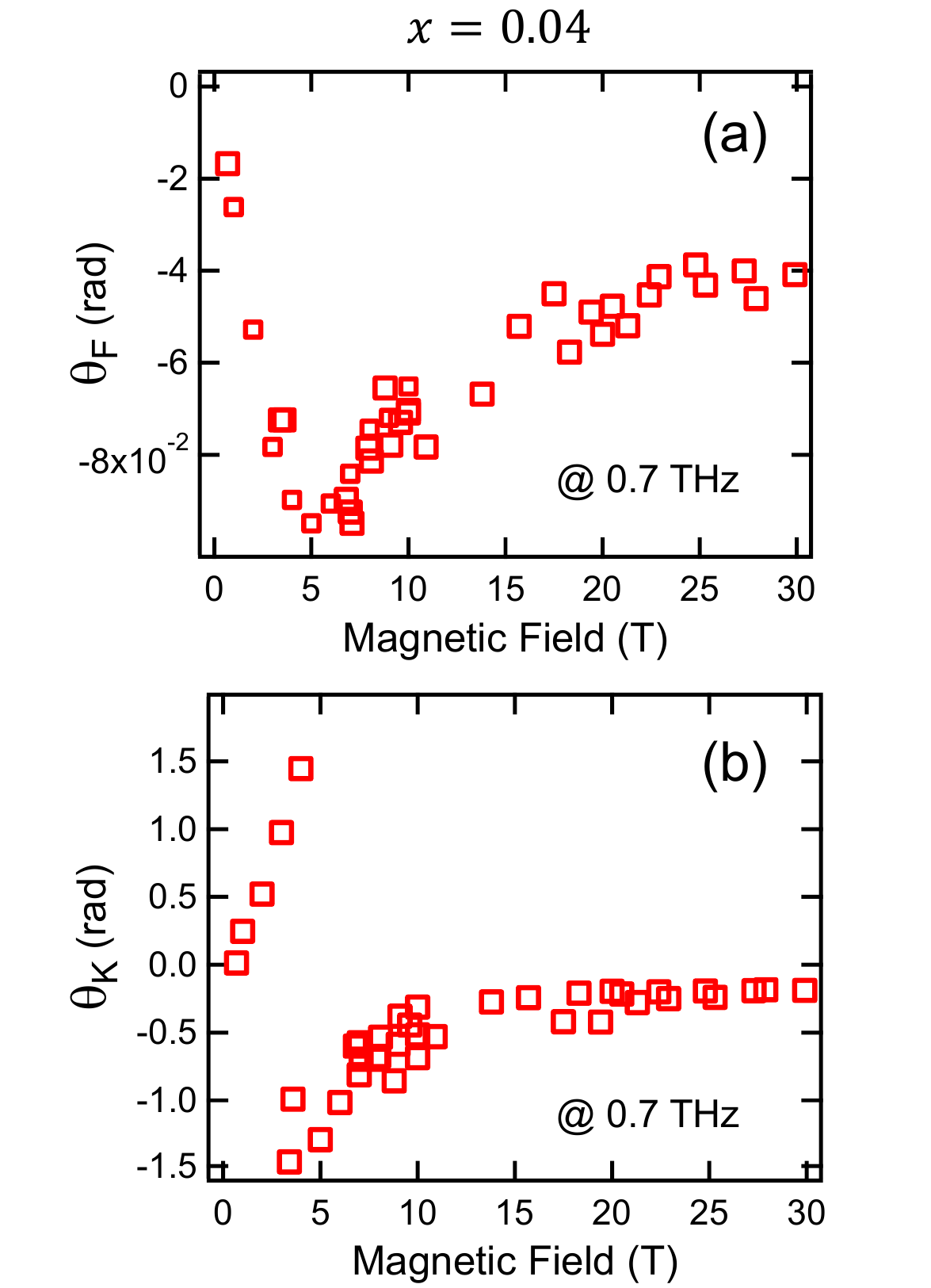}
		\caption{\small (a)~$\theta_\mathrm{F}$ and (b)~$\theta_\mathrm{K}$ versus magnetic field for a fixed THz frequency of 0.7~THz obtained for Sample~2 at $T$ = 21~K in $B$ up to 30~T. }
		\label{Bi096rot30T}
	\end{center}
\end{figure}

\begin{figure*}[htb]
	\begin{center}
		\includegraphics[width=0.6\linewidth]{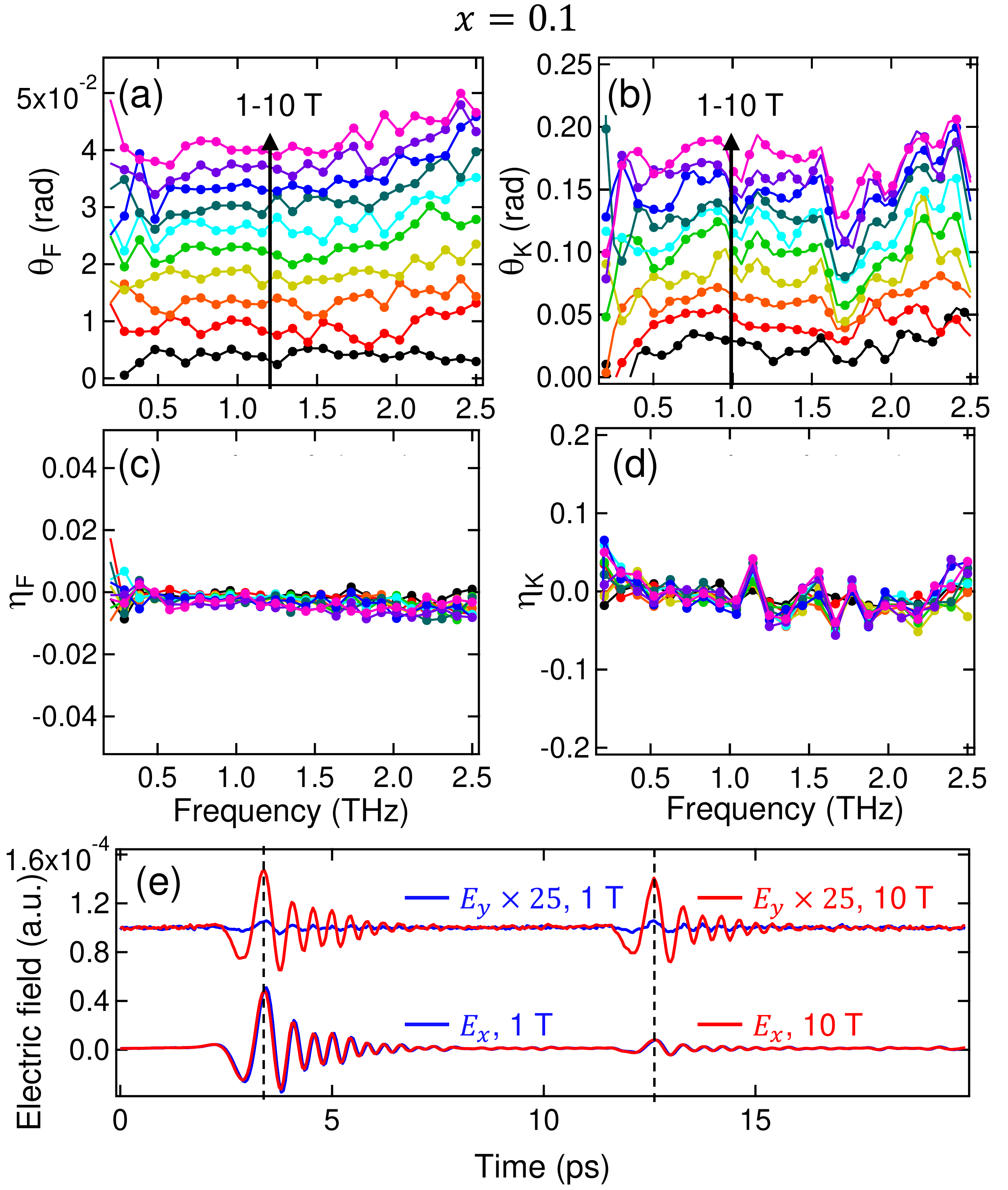}
		\caption{\small Magneto-optical response of Sample 4 (nominal $x$ = 0.1) at $T$ = 2~K in $B$ up to 10~T. (a)~Faraday rotation, (b)~Kerr rotation, (c)~Faraday ellipticity, and (d)~Kerr ellipticity spectra at $B$ from 1 to 10~T are displayed. No curve is intentionally offset. (e)~Time-domain waveforms of $E_x$ and $E_y$ at 1~T and 10~T.  $E_y$ curves are multiplied by a factor of 25 and offset vertically for clarity. Black dashed lines are guides to the eye for identifying the pulse peak positions.}
		\label{TIrotation10T}
	\end{center}
\end{figure*}

\begin{figure*}[htb]
	\begin{center}
		\includegraphics[width=0.6\linewidth]{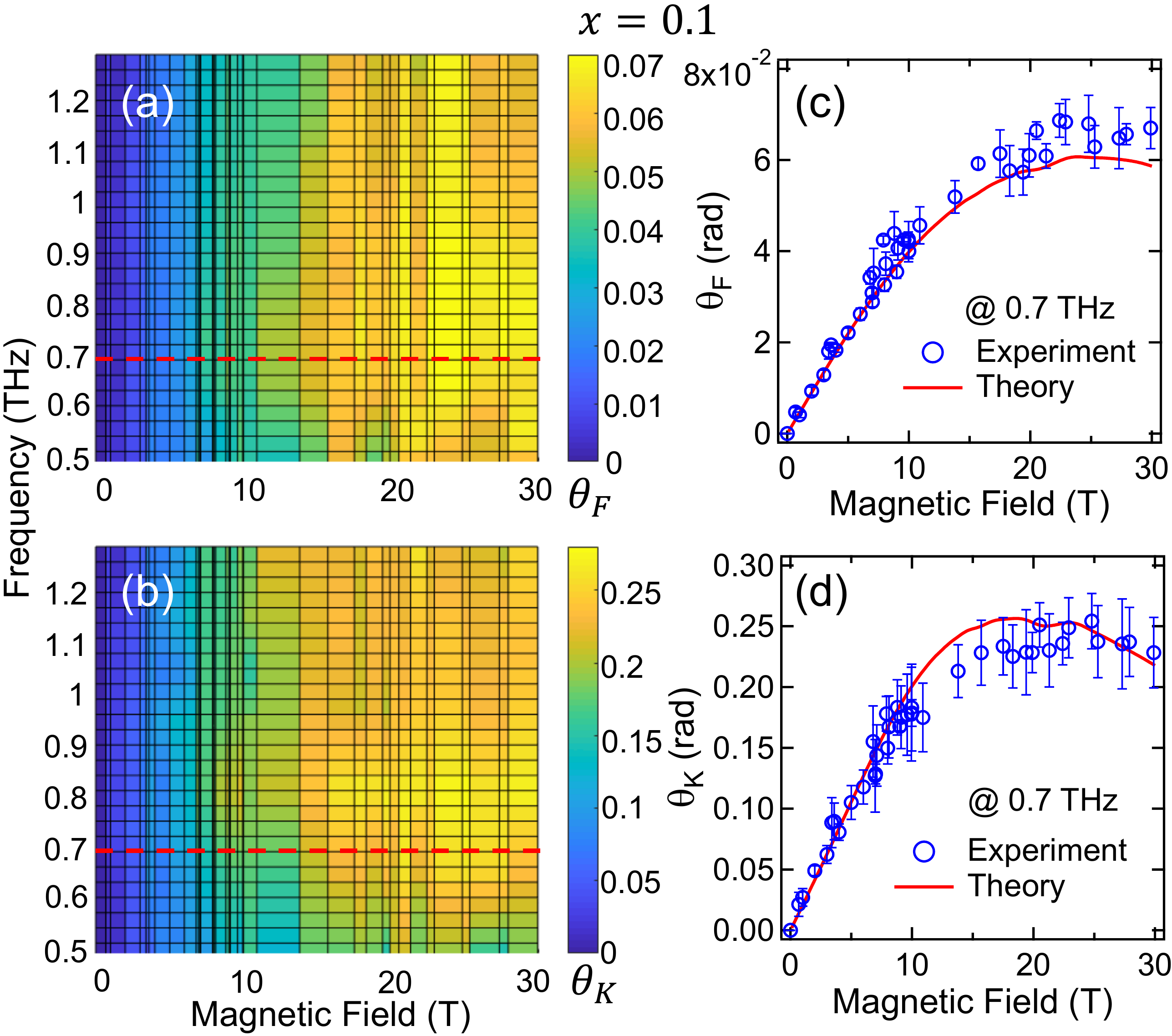}
		\caption{\small Magneto-optical response of Sample 4 (nominal $x$ = 0.1) up to 30~T. (a)~Faraday rotation and (b)~Kerr rotation maps versus THz frequency and magnetic field. (c)~Faraday rotation and (d)~Kerr rotation versus magnetic field at a fixed THz frequency of 0.7~THz, corresponding to the cuts marked by the red dashed lines in (a) and (b). The solid red lines in (c) and (d) are calculated curves using the theoretical model described later in the Discussions section.}
		\label{TIrotation30T}
	\end{center}
\end{figure*}

\begin{figure}[htb]
	\begin{center}
		\includegraphics[scale=0.6]{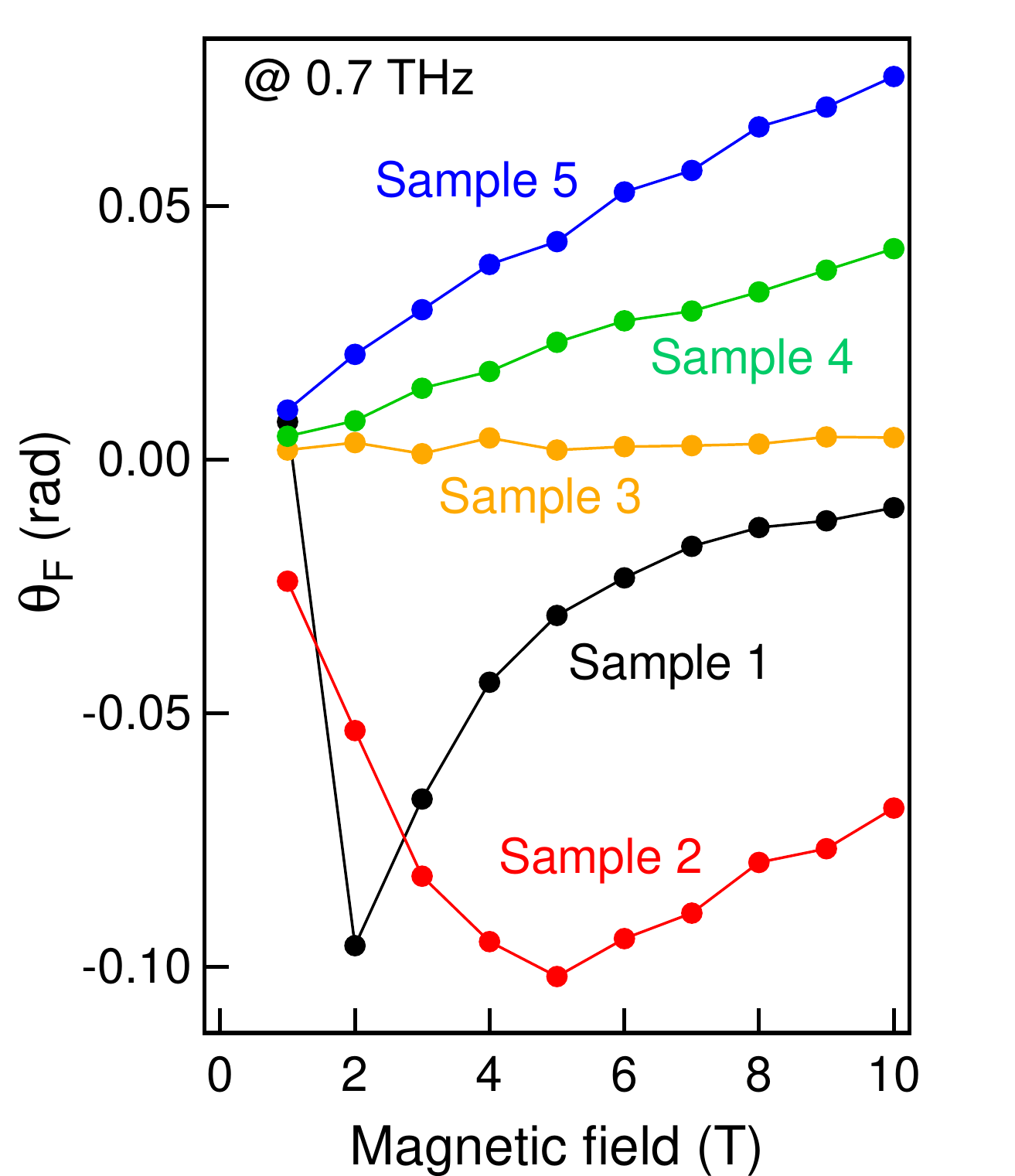}
		\caption{\small Magneto-optical response of all samples. $\theta_\mathrm{F}$ is plotted versus $B$ at $T$ = 2~K for a fixed THz frequency (0.7~THz).  See Table~\ref{BiSbtable} for the $x$ values for the samples.}
		\label{BiSbseriesFaraday}
	\end{center}
\end{figure}

%\newpage

%Here, we present experimental results obtained from the Bi and nominal $\text{Bi}_{0.96}\text{Sb}_{0.04}$ films, which are confirmed to be in the semimetal regime by chemical characterizations.  
Figures~\ref{Biexpt}(a) and (b) display Faraday rotation ($\theta_\mathrm{F}$) and Faraday ellipticity ($\eta_\mathrm{F}$) spectra, respectively, for Sample~1 ($x$ = 0) at $T$ = 2~K in $B$ up to 10~T.  The curves are intentionally offset vertically for clarity, and the zero baselines are shown as dashed colored lines. A resonance feature that shifts higher in frequency with increasing $B$ is clearly observed.  We then calculated the real and imaginary parts of the optical Hall conductivity ($\sigma_{xy}$) of the sample by taking into account the reference signal; see Eq.\,\eqref{sigma_xy}. Figure~\ref{Bisigma} shows the calculated $\sigma_{xy}$ spectra. The signs of the real and imaginary parts of $\sigma_{xy}$ are both opposite to that obtained from the standard 2DEG sample shown in Fig.\,\ref{2DEGHall}. This indicates that holes are the major contributors to the magneto-optical signal.

Figures~\ref{Bi096Faraday}(a) and (b) display $\theta_\mathrm{F}$ and $\eta_\mathrm{F}$ spectra, respectively, for Sample~2 at $T$ = 2~K in $B$ up to 10~T. The major resonance feature that shifts with $B$ in a similar manner to that in Sample~1 is observed, except that the linewidth is broader. This suggests that the major contributors to the magneto-optical signal have the same carrier origin as in the Bi film, and the Sb incorporation reduces the carrier mobility. Figures\,\ref{Bi096Kerr}(a) and (b) show Kerr rotation  ($\theta_\mathrm{K}$) and Kerr ellipticity ($\eta_\mathrm{K}$) spectra, respectively, obtained for the same sample under the same conditions as in Fig.\,\ref{Bi096Faraday}. The resonance feature induces a drastic response in the $\theta_\mathrm{K}$ spectra; an abrupt $\pi$ phase shift of the rotation angle appears at the resonance frequency at each $B$. The $\eta_\mathrm{K}$ spectra, on the other hand, do not show a clear trend, likely due to an insufficient signal-to-noise ratio.

We also performed measurements on Sample~2 in $B$ up to 30~T at $T$ = 21~K, using the single-shot THz detection setup described in Section~\ref{RAMBO}. Figures~\ref{Bi096rot30T}(a) and (b) show, respectively, the extracted $\theta_\mathrm{F}$ and $\theta_\mathrm{K}$ versus $B$ for a fixed THz frequency (0.7~THz). Resonance behavior is again clearly observed in both $\theta_\mathrm{F}$ and $\theta_\mathrm{K}$. $\theta_\mathrm{F}$ shows a dip, while $\theta_\mathrm{K}$ shows a $\pi$ phase shift at the resonance magnetic field. At higher $B$, both quantities decrease in magnitude and tend to zero.

\subsection{Topological insulator sample: Sample 4}
Faraday rotation, Kerr rotation, Faraday ellipticity, and Kerr ellipticity spectra for Sample~4 (nominal $x$ = 0.1 at $T$ = 2~K in $B$ up to 10~T are shown in Figs.\,\ref{TIrotation10T}(a)-(d). No curve is intentionally offset. The Faraday and Kerr rotations are featureless as a function of frequency, but both clearly show an increasing trend as $B$ increases. $\theta_\mathrm{F}$ is smaller than $\theta_\mathrm{K}$ for a given $B$, but its signal-to-noise ratio is higher because the back reflection pulse from which $\theta_\mathrm{K}$ is derived has a smaller amplitude than the main pulse. 

Neither $\eta_\mathrm{F}$ nor $\eta_\mathrm{K}$ shows any signal that can be clearly distinguished from the noise floor for all $B$, suggesting that the carriers in the TI film induce pure rotations without any ellipticity change in the THz probe light. This behavior can be directly observed in the time-domain THz waveform data, as shown in Fig.\,\ref{TIrotation10T}(e). The incident THz pulse is polarized in the $x$ direction. It is clear that the amplitude of the transmitted $E_x$ does not vary much between 1~T and 10~T; at both magnetic fields, the main pulse appears at 3.5~ps, together with a smaller back reflection pulse at 12.5~ps. However, the $E_y$ waveform at 10~T is much larger than that at 1~T, suggesting a much larger polarization rotation; this is consistent with Figs.\,\ref{TIrotation10T}(a) and (b). In addition, as marked by the vertical dashed lines at the pulse peaks, the phase of $E_y$ also matches well with $E_x$ for both the main and back reflection pulses. This suggests that both pulses remain linearly polarized. Only the polarization plane is rotated with respect to the incident beam, and the ellipticity does not change.  

We further performed measurements in $B$ up to 30~T on the same sample at $T$ = 21~K. Figures~\ref{TIrotation30T}(a) and (b) show, respectively, $\theta_\mathrm{F}$ and $\theta_\mathrm{K}$ versus THz frequency and $B$.  The data obtained using the 10-T magnet system (Fig.\,\ref{TIrotation10T}) is also included in Fig.\,\ref{TIrotation30T}, agreeing with the data obtained using the 30~T system. We extracted the evolutions of $\theta_\mathrm{F}$ and $\theta_\mathrm{K}$ with $B$ for a fixed THz frequency (0.7~THz), and the results are plotted in Figs.\,\ref{TIrotation30T}(c) and (d), respectively.  We observe that $\theta_\mathrm{F}$ and $\theta_\mathrm{K}$ increase with increasing $B$ until 15~T, above which both quantities saturate with further increasing $B$.

\subsection{All samples}

%As a final comment, we present the influence of the topological phase transition on the magneto-optical response. 
Figure~\ref{BiSbseriesFaraday} displays $\theta_\mathrm{F}$ versus $B$ at $T$ = 2~K for a fixed THz frequency (0.7~THz) for all samples of $\text{Bi}_{1-x}\text{Sb}_{x}$ films.  As $x$ increases, the system moves from the SM to the TI regime in the phase diagram (Fig.\,\ref{Phasediagram}). We can see a clear trend of magneto-optical response accompanying this transition. From the Bi and $\text{Bi}_{0.96}\text{Sb}_{0.04}$ films in which a bulk hole cyclotron resonance appears as negatively valued dips within 10~T, to the $\text{Bi}_{0.9}\text{Sb}_{0.1}$ film where the low-mobility surface electrons lead to an increasing positive $\theta_\mathrm{F}$ with increasing $B$, the change of $\theta_\mathrm{F}$ curves with increasing $x$ is monotonic and smooth.

%%%%%%%%%%%%%%%%%%%%%%%%%%%%%%%%%%%
%%%%%%%%%%%%%%%%%%%%%%%%%%%%%%%%%%%
\section{Discussion}
%%%%%%%%%%%%%%%%%%%%%%%%%%%%%%%%%%%
%%%%%%%%%%%%%%%%%%%%%%%%%%%%%%%%%%%

\subsection{Theory of magneto-optical response of Bi$_{1-x}$Sb$_x$ in the semimetallic regime}

\begin{figure}[htb]
	\begin{center}
		\includegraphics[scale=0.45]{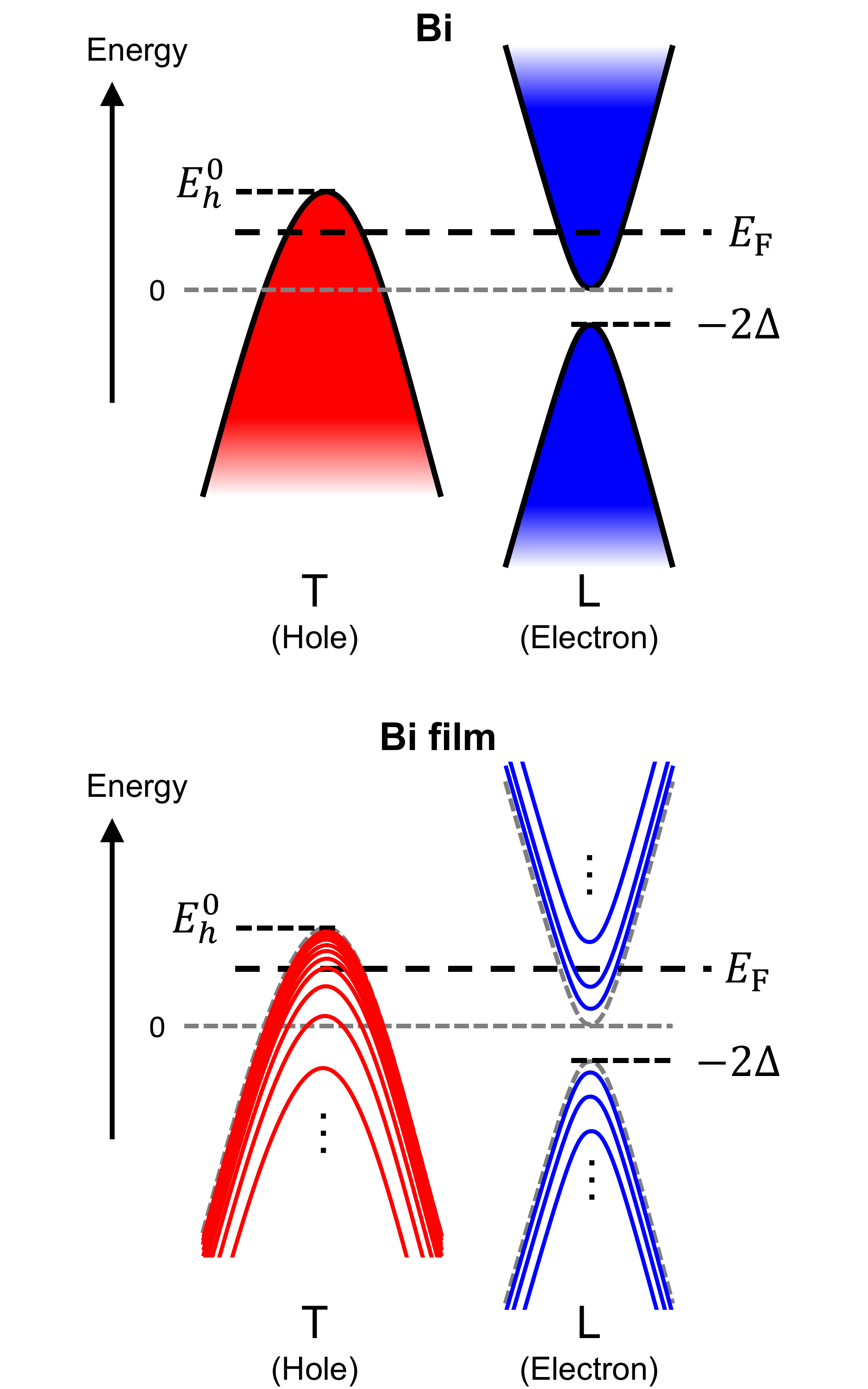}
		\caption{\small Band alignment in the theoretical model used to calculate the magneto-optical response of the Bi film. The bottom panel emphasizes quantum confinement effects in the film case compared to the bulk case shown in the top panel.}
		\label{Bibulkbands}
	\end{center}
\end{figure}

In order to understand the THz magneto-optical response of Bi$_{1-x}$Sb$_x$ films in the SM regime, we developed a detailed theoretical model.  We took into account the bulk bands of Bi, which allowed us to determine the origin of the experimentally observed magneto-optical signal for Sample~1. The reason for choosing the Bi sample instead of $\text{Bi}_{0.96}\text{Sb}_{0.04}$ for this analysis is because the Bi sample showed a much sharper resonance feature.

The bulk band structure we considered is schematically depicted in Fig.\,\ref{Bibulkbands}~\cite{VisseretAl16PRL,ZhuetAl11PRB}. There is an indirect negative band gap between the valence band maximum at the T point and the conduction band minima at the three equivalent L points (later we refer to these as a single point, L). For a $(001)$ Bi film, the hole pocket at the T point has an isotropic in-plane effective mass and a parabolic dispersion relation, while the bands at the L point host Dirac electrons with a hyperbolic dispersion, but a small gap $2\Delta$ exists at the L point.

The Hamiltonian for the T-point holes and the L-point electrons are
\begin{widetext}
	\begin{align}
	\mathcal{H}_h & = E^0_h-\frac{\hbar^2(k_x^2+k_y^2)}{2M_c}-\frac{\hbar^2k_z^2}{2M_z}\\
	\mathcal{H}_e & = 
	\begin{pmatrix}
	\Delta & 0 & i\hbar v_zk_z & i\hbar v(k_x-ik_y)\\
	0 & \Delta & i\hbar v(k_x+ik_y) & -i\hbar v_zk_z\\
	-i\hbar v_zk_z & -i\hbar v(k_x-ik_y) & -\Delta & 0\\
	-i\hbar v(k_x+ik_y) & i\hbar v_zk_z & 0 & -\Delta\\
	\end{pmatrix},
	\end{align}
\end{widetext}
where $E_h^0=38.5\ \text{meV}$ is the T-point band edge offset, $k$ represents the wave vector, $M_c=0.0677m_0$ and $M_z=0.721m_0$ are, respectively, the in-plane and out-of-plane hole effective masses, $m_0$ is the free electron mass, $\Delta=7.65\ \text{meV}$ is half of the L-point gap, and $v$ and $v_z$ are the in-plane and out-of-plane electron Dirac velocities, respectively. Because our film has a finite thickness, we considered the quantum confinement effect on $k_z$ along the growth direction for both the hole and electron pockets. As schematically shown in Fig.\,\ref{Bibulkbands}, many hole subbands described by discrete $k_z$'s with quantum number $N_h$ are above the Fermi energy $E_\text{F}$, while only two electron subbands (described by quantum number $N_e$) are filled at the L point;  the electrons are more strongly influenced by quantum confinement than the holes because of the lighter electron mass.

\begin{figure}[htb]
	\begin{center}
		\includegraphics[scale=0.6]{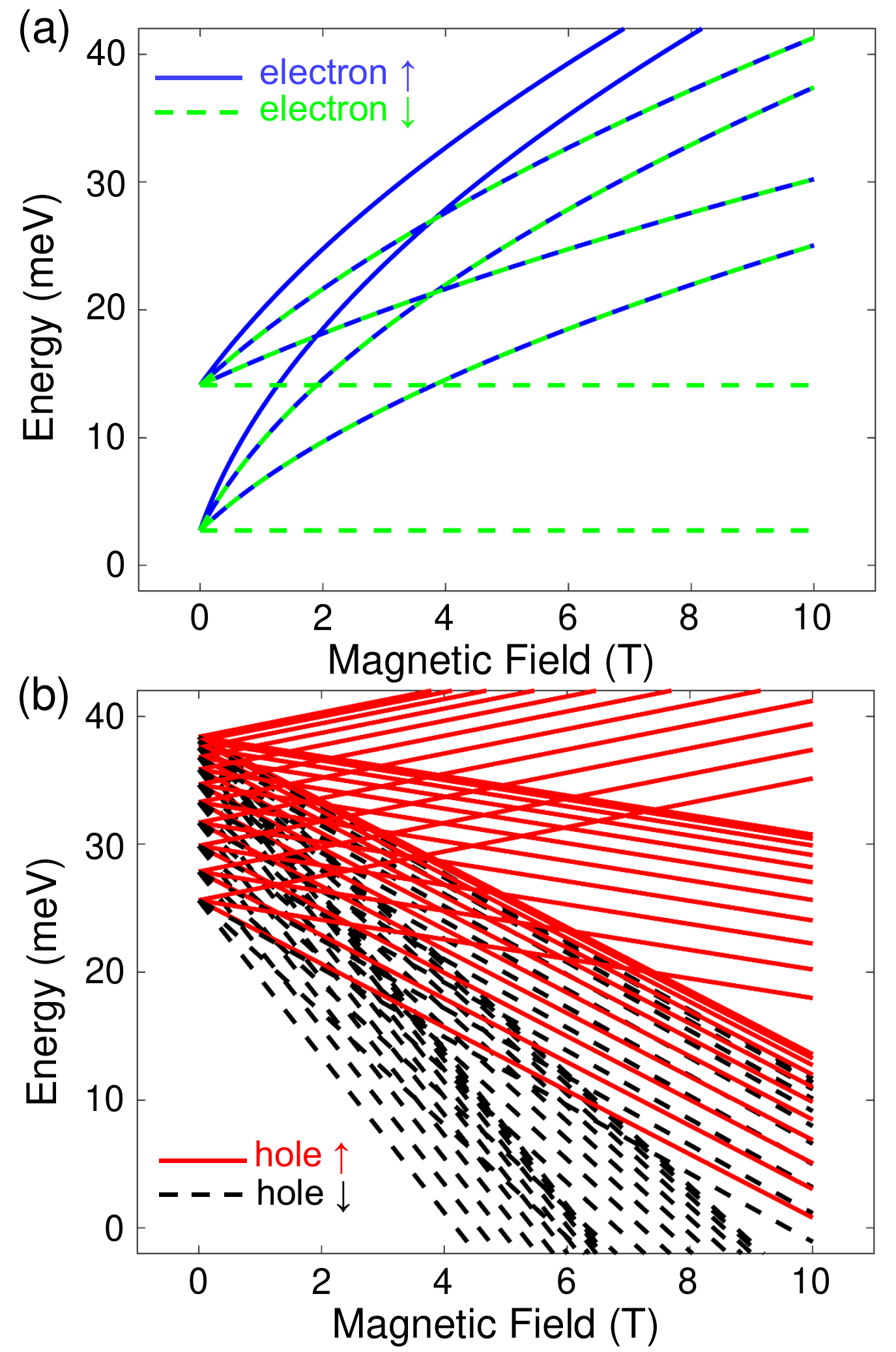}
		\caption{\small Landau-level energies of the (a)~electron and (b)~hole pockets in Bi. The results are displayed up to $N_e=2$ and $n=2$ ($N_h=10$ and $n=2$) for electrons (holes), where $N_e$ and $N_h$ are the subband indices for the electron and hole pocket, respectively, and $n$ represents the Landau-level index.}
		\label{BiLL}
	\end{center}
\end{figure}

The key for the magneto-optical response of carriers is to calculate the optical conductivity tensor given by the Kubo formula:
\begin{align}
\sigma_{\alpha\beta}(\omega) = \frac{i\hbar}{S} \sum_{m,n}
\frac{f_m - f_n}{E_m-E_n}
\frac{\langle \Psi_m|\hat{j}_{\alpha}|\Psi_n\rangle\langle \Psi_n|\hat{j}_{\beta}|\Psi_m\rangle}
{\hbar\omega + E_m - E_n + i\gamma}, \label{Kubo}
\end{align}
where $\alpha$ and $\beta$ take choices between $x$ and $y$, $S$ is the sample area, $f_m$ ($f_n$), $E_m$ ($E_n$), and $|\Psi_m\rangle$ ($|\Psi_n\rangle$) are, respectively, the occupation factor calculated by the Fermi-Dirac distribution function, energy, and eigenfunction of the $m$th ($n$th) eigenstate of the system Hamiltonian, $\hat{j}_{\alpha}$ and $\hat{j}_{\beta}$ are current operators, and $\gamma$ is the scattering rate responsible for transition line broadening. We calculated the conductivity tensors contributed by each pocket and later added all their contributions.

In a magnetic field, the Landau-level eigenenergies for the hole and electron pockets are
\begin{align}
E_{h,n} & = E^h_0-\left(n+\frac{1}{2}\right)\frac{\hbar eB}{M_c}-\frac{\hbar^2k_z^2}{2M_z}+s_hg\mu_\text{B}B\\
E_{e,n} & = \sqrt{\Delta^2+2\Delta\left(n+\frac{1}{2}+s_e\right)\frac{\hbar eB}{m_c}+\frac{\hbar^2k_z^2}{2m_z}},
\end{align}
where $s_h=\pm1/2$ and $s_e=\pm1/2$ are, respectively, the hole and electron spin quantum number, $n$ represents the Landau-level index, $g=62.6$ is the hole $g$ factor, $\mu_\text{B}$ is the Bohr magneton, $m_c$ and $m_z$ are, respectively, the in-plane and out-of-plane electron effective masses. The calculated spin-resolved Landau-level energies for electrons and holes are plotted in Fig.\,\ref{BiLL}(a) and (b), respectively. The results are displayed up to $N_e=2$ and $n=2$ ($N_h=10$ and $n=2$) for electrons (holes), where $N_e$ and $N_h$ are the electron and hole subband index, respectively. 

Then we calculated the matrix elements of the current operators, $\langle \Psi_m|\hat{j}_{\alpha}|\Psi_n\rangle$, for both electron and hole pockets. For the electron pocket, the situation is complicated because the eigenspinor of the Dirac Hamiltonian above takes separate forms for spin-up and spin-down electrons
\begin{align}
\Psi_{n,\downarrow} & = \begin{pmatrix}
A_{n,\downarrow}|n-1\rangle \\
B_{n,\downarrow}|n\rangle \\
C_{n,\downarrow}|n-1\rangle \\
D_{n,\downarrow}|n\rangle \\	
\end{pmatrix}\\
\Psi_{n,\uparrow} & = \begin{pmatrix}
A_{n,\uparrow}|n\rangle \\
B_{n,\uparrow}|n+1\rangle \\
C_{n,\uparrow}|n\rangle \\
D_{n,\uparrow}|n+1\rangle \\	
\end{pmatrix},
\end{align}
where $\downarrow$ and $\uparrow$ denote electron spin, and $A_{n,\downarrow}$, $B_{n,\downarrow}$, $C_{n,\downarrow}$, $D_{n,\downarrow}$, $A_{n,\uparrow}$, $B_{n,\uparrow}$, $C_{n,\uparrow}$, and $D_{n,\uparrow}$ are the coefficients to be determined by the eigenvalue equation. Therefore, there are eight matrix elements of the current operator describing possible transitions between Landau levels across different spin channels. These are $\langle \Psi_{m,\downarrow}|\hat{j}_{x}|\Psi_{n,\downarrow}\rangle$, $\langle \Psi_{m,\uparrow}|\hat{j}_{x}|\Psi_{n,\uparrow}\rangle$, $\langle \Psi_{m,\downarrow}|\hat{j}_{x}|\Psi_{n,\uparrow}\rangle$, $\langle \Psi_{m,\uparrow}|\hat{j}_{x}|\Psi_{n,\downarrow}\rangle$, $\langle \Psi_{m,\downarrow}|\hat{j}_{y}|\Psi_{n,\downarrow}\rangle$, $\langle \Psi_{m,\uparrow}|\hat{j}_{y}|\Psi_{n,\uparrow}\rangle$, $\langle \Psi_{m,\downarrow}|\hat{j}_{y}|\Psi_{n,\uparrow}\rangle$, and $\langle \Psi_{m,\uparrow}|\hat{j}_{y}|\Psi_{n,\downarrow}\rangle$.

\begin{figure}[htb]
	\begin{center}
		\includegraphics[scale=0.45]{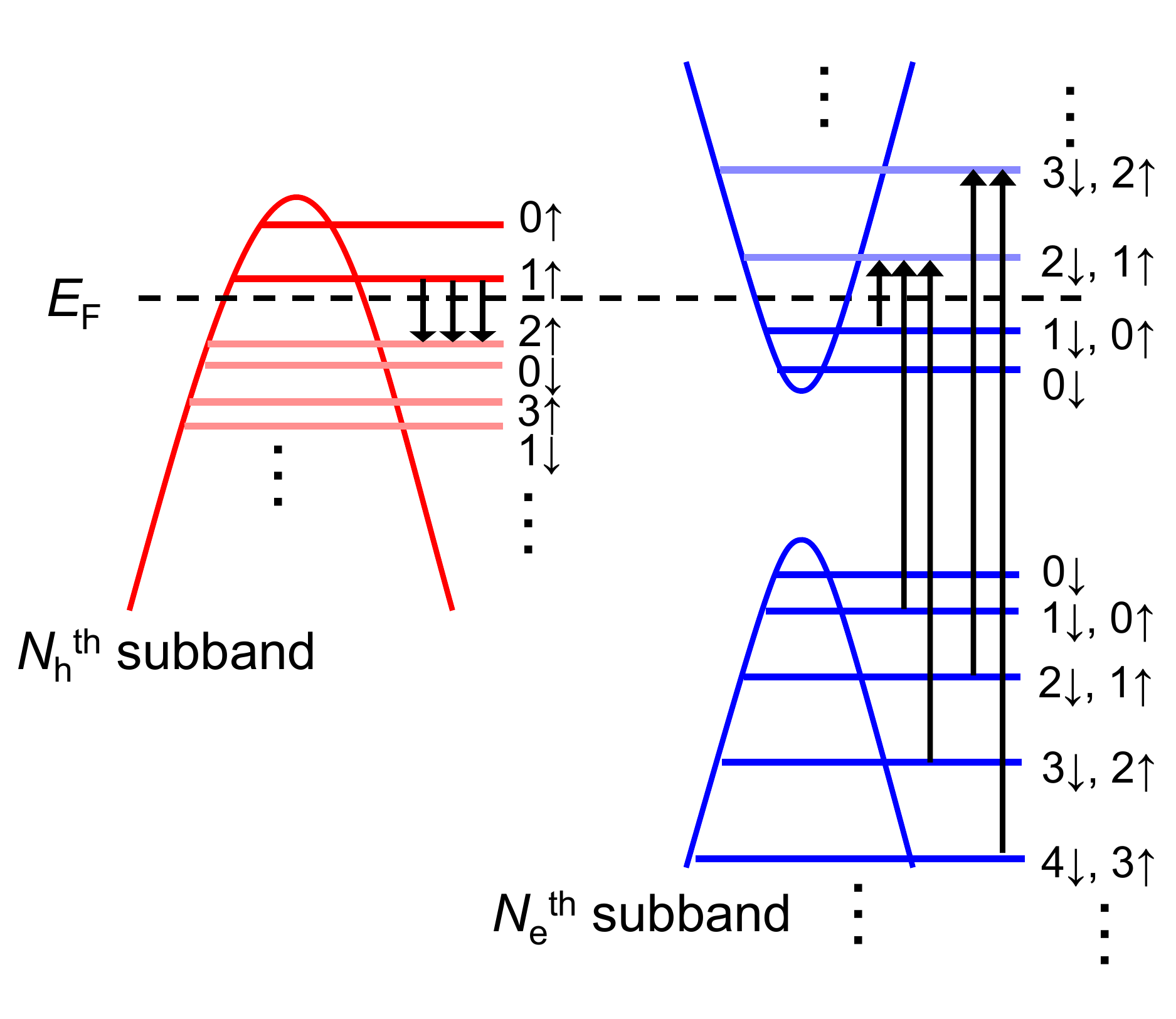}
		\caption{\small Transitions considered for the hole pocket within the $N_h$th subband and the electron pocket within the $N_e$th subband in Bi. States are labeled by their Landau-level index and spin polarization. The black arrows indicate allowed magneto-optical transitions.}
		\label{BiMOtransitions}
	\end{center}
\end{figure}

We calculated optical conductivity tensors by plugging the matrix elements of the current operator above into the Kubo formula. The total conductivity of each carrier pocket was obtained by summing up transitions between all possible states. Each state has five indices, i.e., the Landau-level index, spin index, band index describing valence or conduction band, subband index, and the pocket index due to the existence of multiple equivalent pockets. As an example, we show the transitions considered for the hole pocket within the $N_h$th subband and the electron pocket within the $N_e$th subband in Fig.\,\ref{BiMOtransitions}. The hole pocket contains only the spin-conserving inter-Landau-level transitions. The electron pocket contains intraband inter-Landau-level transitions and interband magneto-optical transitions across different spin channels and Landau-level indices. Note that intersubband transitions are not allowed in either the hole or electron pocket due to the polarization selection rule. 

\begin{figure}[htb]
	\begin{center}
		\includegraphics[scale=0.4]{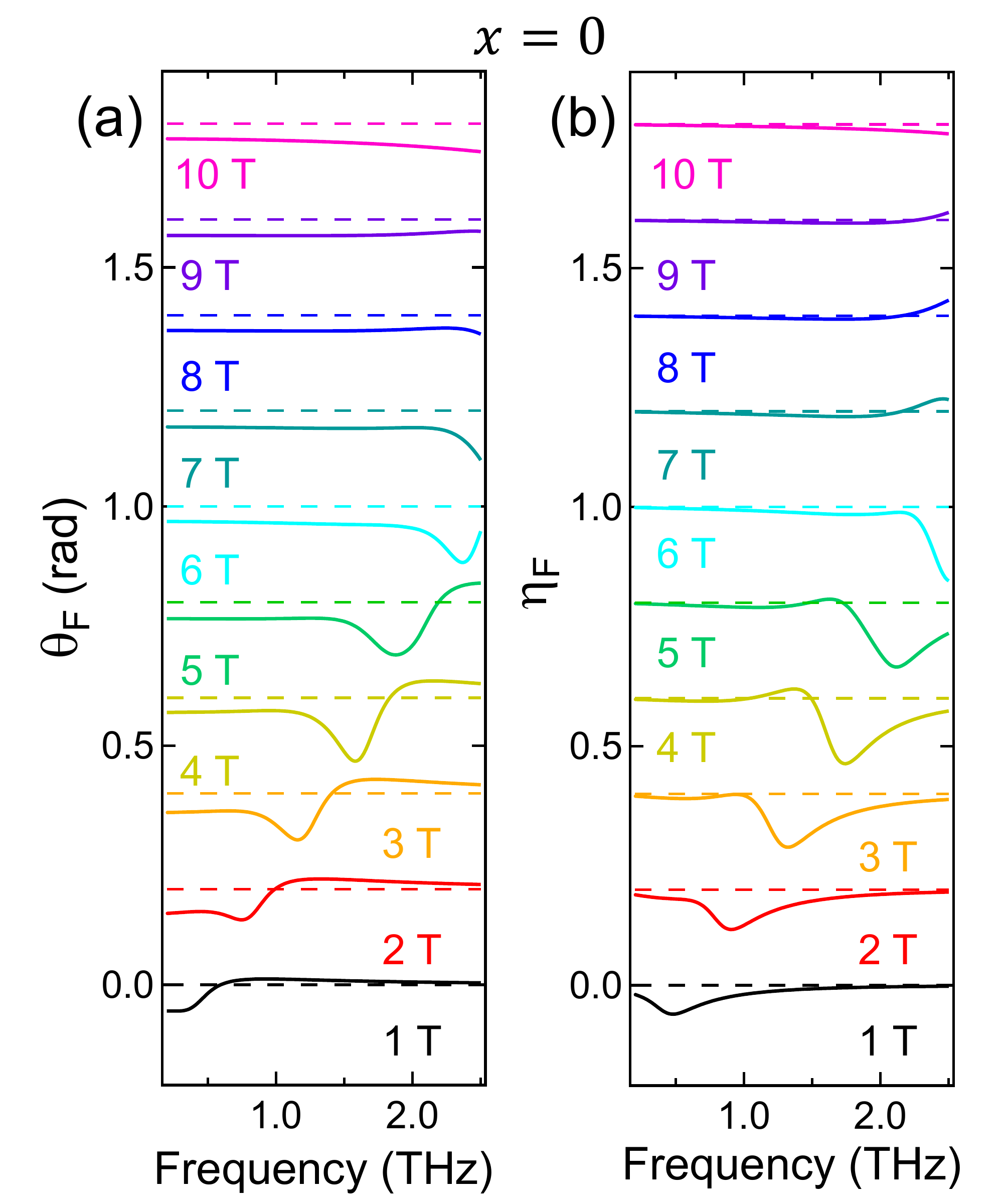}
		\caption{\small Calculated (a)~Faraday rotation and (b)~Faraday ellipticity spectra for the Bi film at $T$ = 2~K in $B$ up to 10~T.  Curves at different $B$ are vertically offset intentionally for clarity, and zero baselines are marked by dashed lines.}
		\label{BicalFaraday}
	\end{center}
\end{figure}

Finally, we summed up the conductivities of the electron and hole pockets to obtain the total optical conductivity tensor of the Bi film. We adjusted three parameters to try to match the theoretically calculated magneto-optical response with the experiments. They are the Fermi energy $E_\text{F}$, the scattering rate of holes at the T point $\gamma^\text{T}_h$, and the scattering rate of electrons at the L point $\gamma^\text{L}_e$. The optimized parameters that achieve the best agreement between theory and experiment are $E_\text{\text{F}}=24\ \text{meV}$, $\gamma^\text{T}_h=0.75\ \text{meV}$, and $\gamma^\text{L}_e=0.9\ \text{meV}$. The calculated $\theta_\mathrm{F}$ and $\eta_\mathrm{F}$ spectra for the Bi film sample at 2~K up to 10~T are shown in Fig.\,\ref{BicalFaraday}; the curves show good agreement with the experimental data in Fig.\,\ref{Biexpt}.

\begin{figure}[htb]
	\begin{center}
		\includegraphics[scale=0.4]{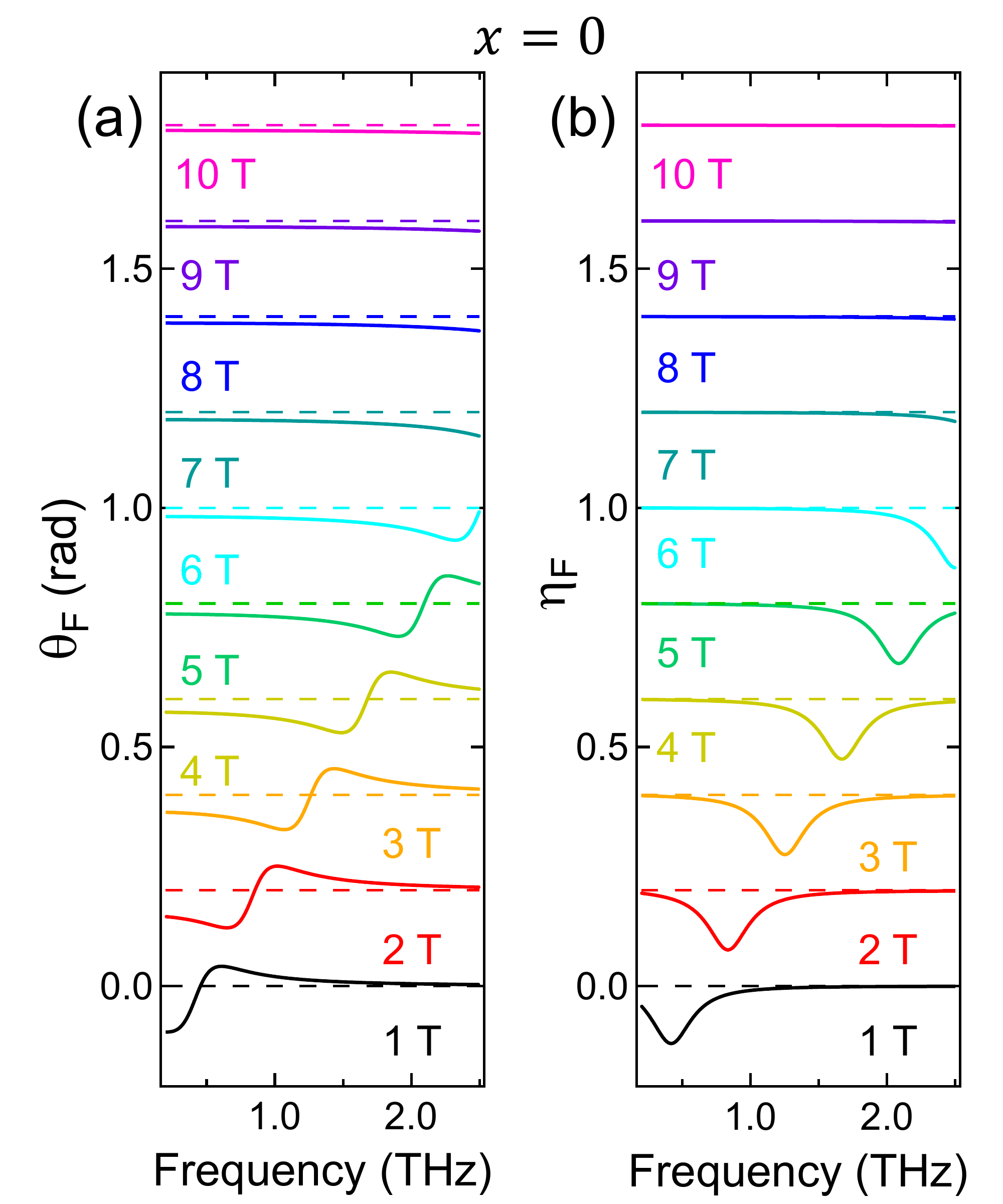}
		\caption{\small Calculated (a)~Faraday rotation and (b)~Faraday ellipticity spectra for the Bi film at $T$ = 2~K in $B$ up to 10~T by considering only the hole pocket. Curves at different $B$ are vertically offset intentionally for clarity, and zero baselines are marked by dashed lines.}
		\label{BicalFaradayHole}
	\end{center}
\end{figure}

We provide some comments on the optimized parameters. First, the extracted Fermi energy $E_\text{F}=24\ \text{meV}$ is lower than the value (28 meV) used in previous studies~\cite{VisseretAl16PRL,ZhuetAl11PRB}. The reason can be that the bands are modified compared to the true bulk Bi due to the band quantizations resulting from the finite thickness of the film. Second, $\gamma^\text{T}_h=0.75\ \text{meV}$ and $\gamma^\text{L}_e=0.9\ \text{meV}$ are both much smaller than the bandwidth of our THz setup. This therefore allows cyclotron resonance signals to be observed. From the calculated results in Fig.\,\ref{BicalFaraday}, we found that the dip feature that moves to higher frequency with increasing magnetic field is mainly due to hole cyclotron resonance. The electron cyclotron resonance feature, on the other hand, appears very close in frequency with the hole cyclotron resonance feature, but it is inhomogeneously broadened because cyclotron transition frequencies are different for different electron subbands. The effect of electron cyclotron resonance can still be identified in the calculated results in Fig.\,\ref{BicalFaraday}. We plotted the $\theta_\mathrm{F}$ and $\eta_\mathrm{F}$ spectra calculated by only taking into account the hole pocket in Fig.\,\ref{BicalFaradayHole}. It can be easily seen that the hole pocket alone cannot explain the asymmetric lineshapes of $\theta_\mathrm{F}$ and $\eta_\mathrm{F}$ in Fig.\,\ref{BicalFaraday} and in the experimental data in Fig.\,\ref{Biexpt}.

\subsection{Theory of magneto-optical response of Bi$_{1-x}$Sb$_x$ in the topological insulator regime}

\begin{figure}[htb]
	\begin{center}
		\includegraphics[scale=0.5]{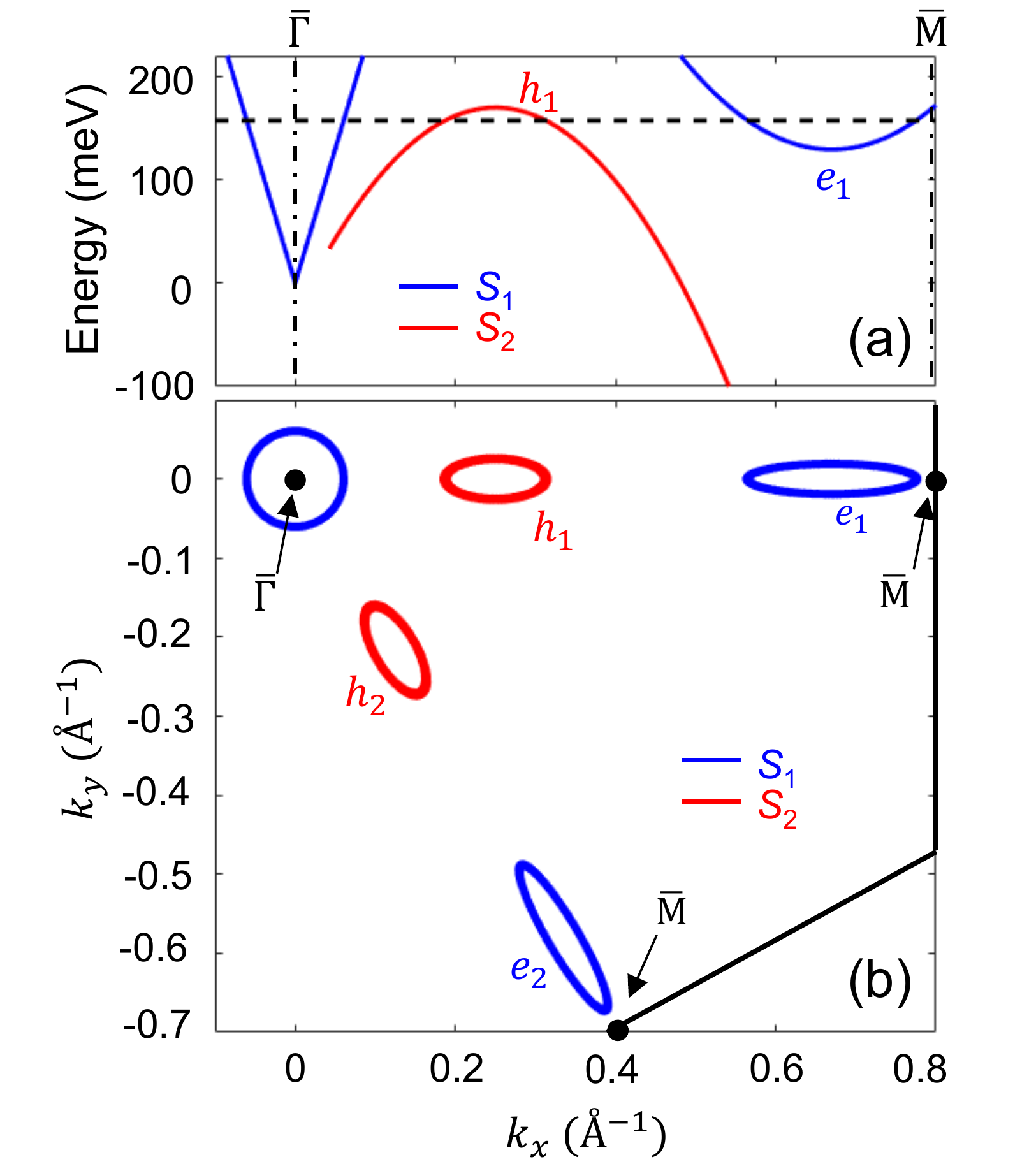}
		\caption{\small Surface band model used in the theoretical analysis for Bi$_{1-x}$Sb$_x$ in the topological insulator regime. (a)~Band dispersions along the $\bar{\Gamma}$-$\bar{\text{M}}$ line. The black dashed line marks the position of the Fermi energy. The black dashed-dotted lines mark the positions of the $\bar{\Gamma}$ and $\bar{\text{M}}$ points. (b)~Fermi surface within one of the quadrants of the first Brillouin zone. Solid black lines are the boundaries of the first Brillouin zone.}
		\label{surfbandFermi}
	\end{center}
\end{figure}

We used a surface band model to analyze the experimental observations made in Sample~4, a TI sample~\cite{ZhangetAl09PRB,BeniaetAl15PRB}; see Fig.\,\ref{surfbandFermi}. The $(001)$ surface of Bi$_{0.9}$Sb$_{0.1}$ has two spin-polarized surface bands, $S_1$ and $S_2$. The dispersion relation along the $\bar{\Gamma}$-$\bar{\text{M}}$ line has the following features. The bottom of the electron pocket formed by the $S_1$ band is located at the $\bar{\Gamma}$ point; the band has a linear dispersion, and its spin texture is similar to that of typical TI surface states with spin-momentum locking. Away from the $\bar{\Gamma}$ point, the $S_1$ band bends up and then down to intersect with the Fermi surface again, forming an anisotropic electron pocket near the $\bar{\text{M}}$ point (but not enclosing it); later we refer to it as the $\bar{\text{M}}$-point electron pocket. On the other hand, the $S_2$ surface band forms an anisotropic hole pocket in the middle of the $\bar{\Gamma}$-$\bar{\text{M}}$ line. Both the $\bar{\text{M}}$-point electron pocket and the hole pocket have six replicas in the first Brillouin zone, which are related to each other by a six-fold rotational symmetry with respect to the $\bar{\Gamma}$ point. We denote the $\bar{\text{M}}$-point electron pocket and the hole point along the $\bar{\Gamma}$-$\bar{\text{M}}$ line to be $e1$ and $h1$, respectively, and their replicas as $e2\cdots,e6$, and $h2\cdots,h6$.

The model Hamiltonians for the $\bar{\Gamma}$-point electron pocket, the $h1$ hole pocket, and the $e1$ electron pocket are
\begin{align}
\mathcal{H}^{\bar{\Gamma}}_e & = v_\text{D} \bm{\sigma} \cdot \bm{p}\\
\mathcal{H}_{h1} & = E^0_h - \frac{(p_x-p_{h0})^2}{2m^h_x} - \frac{p_y^2}{2m^h_y} \\
\mathcal{H}^{\bar{\text{M}}}_{e1} & = E^0_{e} + \frac{(p_x-p_{e0})^2}{2m^e_x} + \frac{p_y^2}{2m^e_y},
\end{align} 
where $v_\text{D}=2600\ \text{meV\AA}$ is the Dirac velocity, $p$ represents the carrier momentum, $E^0_h=170\ \text{meV}$ and $E^0_e=129.4\ \text{meV}$ are the band-edge energies of the hole pocket and the $\bar{\text{M}}$-point electron pocket, respectively, measured from the Dirac point, $p_{h0}=\hbar\times0.25\ \text{meV/\AA}$ and $p_{e0}=\hbar\times0.67\ \text{meV/\AA}$ are the displacements of band centers of the $h1$ and $e1$ pockets, respectively, and in-plane anisotropic effective masses can be described as $m^h_x=1.2m_0$, $m^h_y=0.2m_0$, $m^e_x=1.5m_0$, and $m^e_y=0.05m_0$, where $m_0$ is the free electron mass. The surface band dispersions calculated by applying these parameters are shown in Fig.\,\ref{surfbandFermi}(a). The Fermi surfaces of the pockets within one quadrant of the first Brillouin zone are shown in Fig.\,\ref{surfbandFermi}(b); the Fermi level is obtained from a fit to the Fermi surface of the $\bar{\Gamma}$-point electron pocket measured by ARPES, as shown in Fig.\,\ref{Bi090electronic}(a).

\begin{figure}[htb]
	\begin{center}
		\includegraphics[scale=0.55]{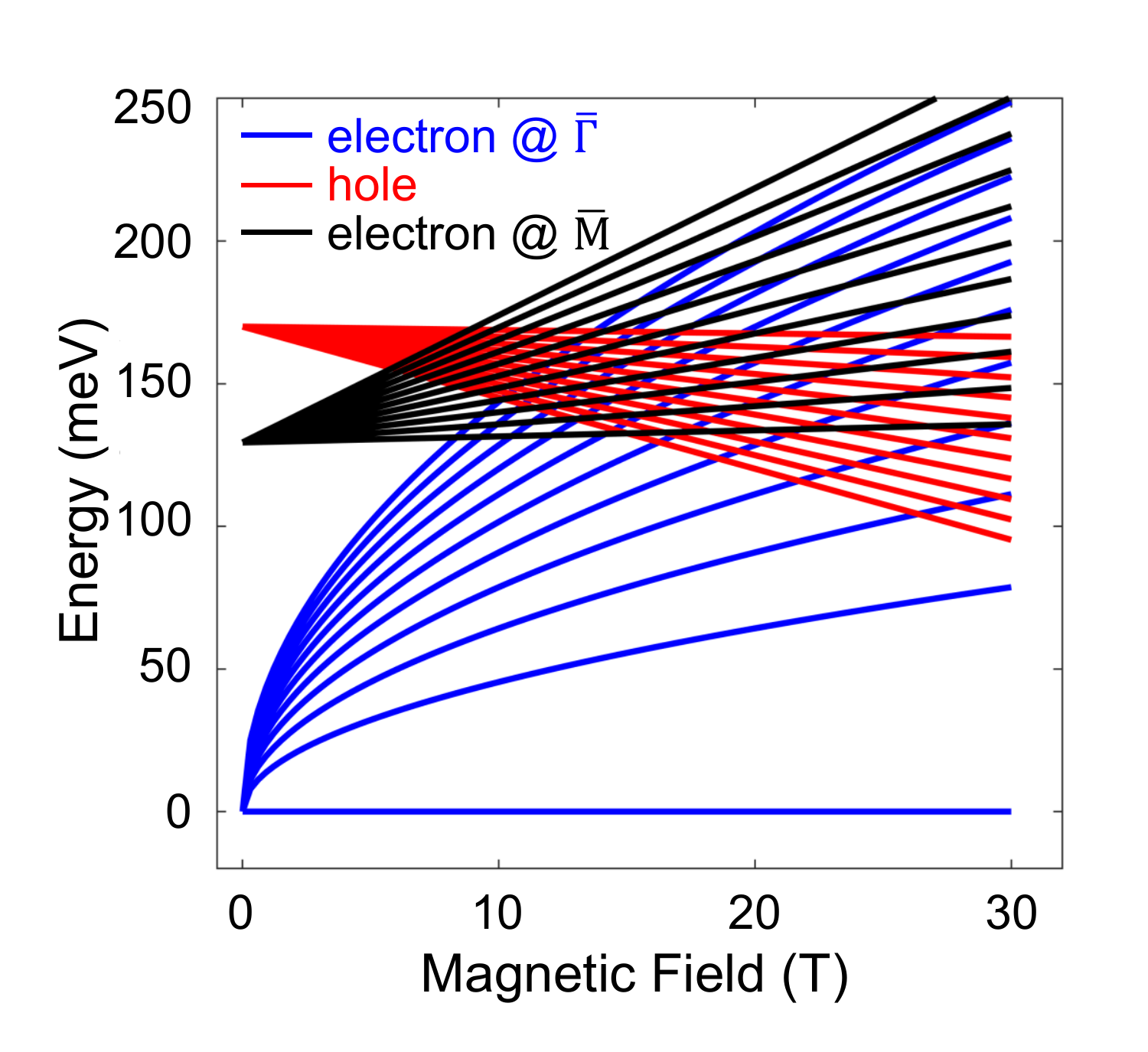}
		\caption{\small Landau-level energies of the three carrier pockets calculated for Bi$_{0.9}$Sb$_{0.1}$ within the surface band model. The $\bar{\Gamma}$ and $\bar{\text{M}}$ points are depicted in Fig.\,\ref{surfbandFermi}.}
		\label{TIsurfaceLL}
	\end{center}
\end{figure}

First, in a magnetic field, the Hamiltonian for the $\bar{\Gamma}$-point electron pocket is given by
\begin{align}
\mathcal{H}_e & = 
\begin{pmatrix}
0 &  \dfrac{\sqrt{2}\hbar v_\text{D}}{\ell_B} \hat{a}\\
\dfrac{\sqrt{2}\hbar v_\text{D}}{\ell_B} \hat{a}^\dagger & 0
\end{pmatrix},
\end{align}
where $l_B=\sqrt{\hbar/|e|B}$ is the magnetic length, and $\hat{a}$ ($\hat{a}^\dagger$) is the Landau-level raising (lowering) operator. Its eigenstate spinor takes the form
\begin{align}
|\Psi_n\rangle = 
\begin{pmatrix}
A_n |n-1\rangle \\ B_n |n\rangle
\end{pmatrix},
\end{align}
where $n$ is the Landau-level index, and $A_n$ and $B_n$ are coefficients to be determined by the following eigenvalue problem:
\begin{align} 
\begin{pmatrix}
0 &  \dfrac{\sqrt{2}\hbar v_\text{D}}{\ell_B} \sqrt{n}\\
\dfrac{\sqrt{2}\hbar v_\text{D}}{\ell_B} \sqrt{n} & 0
\end{pmatrix}
\begin{pmatrix}
A_n\\ B_n 
\end{pmatrix}
= E_n 
\begin{pmatrix}
A_n\\ B_n 
\end{pmatrix}.
\end{align}
The Landau-level eigenenergies are obtained as
\begin{align}
E_n=\text{sgn}(n)\sqrt{\frac{2\hbar^2v_\text{D}^2}{\ell^2}|n|}. \ \ \ \, n=0,1,2,\cdots
\end{align}
See Fig.\,\ref{TIsurfaceLL} for the calculated eigenenergies versus magnetic field up to the $n=10$th Landau level. The corresponding coefficients are found to be
\begin{align}
A_n &= \text{sgn}(n) \sqrt{\frac{1}{2}} \\
B_n &= \sqrt{\frac{1}{2}}.
\end{align}
Given that the current operator is given by 
\begin{align}
\hat{\bm{j}} = \frac{ie}{\hbar}[\mathcal{H}_e, \hat{\bm{x}}],
\end{align}
its matrix elements can be calculated as
\begin{align}
\langle \Psi_m|\hat{j}_{x}|\Psi_n\rangle &= ev_\text{D} (A_mB_n\delta_{m-1,n}+B_mA_n\delta_{m,n-1}) \\
\langle \Psi_m|\hat{j}_{y}|\Psi_n\rangle &= iev_\text{D} (-A_mB_n\delta_{m-1,n}+B_mA_n\delta_{m,n-1}),
\end{align}
where $\delta_{m-1,n}$ and $\delta_{m,n-1}$ are Kronecker's $\delta$. By plugging the matrix elements above into the Kubo formula, we obtained the longitudinal and Hall conductivities for the $\bar{\Gamma}$-point electron pocket
\begin{widetext}
	\begin{align}
	\sigma^{(e,\bar{\Gamma})}_{xx} &= \frac{i\hbar e^2v_\text{D}^2}{S} \sum_m
	\frac{f_m - f_{m+1}}{E_m-E_{m+1}} A^2_{m+1}B^2_{m}
	\left(\frac{1}{\hbar\omega + E_m - E_{m+1} + i\gamma^{\bar{\Gamma}}_e}+\frac{1}{\hbar\omega + E_{m+1} - E_{m} + i\gamma^{\bar{\Gamma}}_e}\right) \\
	\sigma^{(e,\bar{\Gamma})}_{xy} &= \frac{\hbar e^2v_\text{D}^2}{S} \sum_m
	\frac{f_m - f_{m+1}}{E_m-E_{m+1}} A^2_{m+1}B^2_{m}
	\left(\frac{1}{\hbar\omega + E_m - E_{m+1} + i\gamma^{\bar{\Gamma}}_e}-\frac{1}{\hbar\omega + E_{m+1} - E_{m} + i\gamma^{\bar{\Gamma}}_e}\right),
	\end{align}
\end{widetext}
where $\gamma^{\bar{\Gamma}}_e$ represents the scattering rate of electrons in the $\bar{\Gamma}$-point electron pocket.

Second, the Hamiltonian for the hole pocket in a magnetic field is
\begin{align}
\mathcal{H}_{h1} & = E^0_{h} - \hbar\omega^h_c(\hat{a}^{\dagger}\hat{a}+1/2),
\end{align}
where $\omega^h_c$ is the hole cyclotron frequency. The form is simple because the pocket can be assumed to have a parabolic band dispersion. Landau-level energies depend linearly on the magnetic field; see Fig.\,\ref{TIsurfaceLL}.

Regarding the treatment of effective mass anisotropy for the hole pocket, it has been found that the system can be transformed into an isotropic model, which is easier for calculation, by performing a scaling procedure on the spatial metric~\cite{VisseretAl16PRL}. We scaled the current operator as
\begin{align}
\hat{\tilde{j}}_{x} &=\eta^{-1} \hat{j}_{x} \\
\hat{\tilde{j}}_{y} &=\eta  \hat{j}_{y},
\end{align}
where $\eta=(m^h_x/m^h_y)^{1/4}$ is the scaling factor. Then the matrix elements of $\hat{\tilde{j}}_{\alpha}$ can be written as
%\begin{widetext}
\begin{align}
\langle m|\hat{\tilde{j}}_{x}|n\rangle &= -\frac{e\hbar}{\sqrt{2}M\ell_B}\langle m|\hat{a}+\hat{a}^{\dagger}|n\rangle\\
&=-\frac{e\hbar}{\sqrt{2}M\ell_B} (\sqrt{n}\delta_{m,n-1}+\sqrt{n+1}\delta_{m,n+1}) \\
\langle m|\hat{\tilde{j}}_{y}|n\rangle &= \frac{ie\hbar}{\sqrt{2}M\ell_B}\langle m|\hat{a}^{\dagger}-\hat{a}|n\rangle\\
&=\frac{ie\hbar}{\sqrt{2}M\ell_B} (\sqrt{n+1}\delta_{m,n+1}-\sqrt{n}\delta_{m,n-1}),
\end{align}
where $M=\sqrt{m^h_x m^h_y}$. Substituting the matrix elements above into the Kubo formula gives the scaled conductivity tensor
\begin{widetext}
	\begin{align}
	\tilde{\sigma}^{(h1)}_{xx}(\omega) &= \frac{ie^2\hbar^3}{2M^2\ell_B^2S}\sum_{m}
	\frac{f_m - f_{m+1}}{E_m-E_{m+1}}
	\left( \frac{m+1}{\hbar\omega + E_{m} - E_{m+1} + i\gamma_h} + \frac{m+1}{\hbar\omega + E_{m+1} - E_m + i\gamma_h}  \right) \\
	\tilde{\sigma}^{(h1)}_{xy}(\omega) &= \frac{e^2\hbar^3}{2M^2\ell_B^2S}\sum_{m}
	\frac{f_m - f_{m+1}}{E_m-E_{m+1}}
	\left( \frac{m+1}{\hbar\omega + E_{m} - E_{m+1} + i\gamma_h} - \frac{m+1}{\hbar\omega + E_{m+1} - E_m + i\gamma_h}  \right),
	\end{align}
\end{widetext}
where $\gamma_h$ represents the scattering rate of the hole pocket. The actual conductivity tensor is related to the scaled conductivity tensor by
\begin{align}
\sigma^{(h1)}_{xx}(\omega) &= \eta^2	\tilde{\sigma}^{(h1)}_{xx}(\omega) \\
\sigma^{(h1)}_{yy}(\omega) &= \eta^{-2}	\tilde{\sigma}^{(h1)}_{yy}(\omega) \\
\sigma^{(h1)}_{xy}(\omega) &= \tilde{\sigma}^{(h1)}_{xy}(\omega) \\
\sigma^{(h1)}_{yx}(\omega) &=  -\sigma^{(h1)}_{xy}(\omega) = -\tilde{\sigma}^{(h1)}_{xy}(\omega) = \tilde{\sigma}^{(h1)}_{yx}(\omega).
\end{align}

The optical conductivity tensor for the rest of hole pockets $h2,\ h3,\ \cdots,\ h6$ can be obtained by six-fold rotations. The total conductivity as a sum of contributions from all hole pockets is obtained as
\begin{align}
\sigma^{(h)}(\omega) &= \sum_{i=1}^6	\sigma^{(hi)}(\omega) \\
&=3
\begin{pmatrix}
\sigma^{(h1)}_{xx}(\omega)+\sigma^{(h1)}_{yy}(\omega) & 2\sigma^{(h1)}_{xy}(\omega) \\
- 2\sigma^{(h1)}_{xy}(\omega) & \sigma^{(h1)}_{xx}(\omega)+\sigma^{(h1)}_{yy}(\omega) 
\end{pmatrix}.
\end{align}

\begin{figure}[htb]
	\begin{center}
		\includegraphics[scale=0.45]{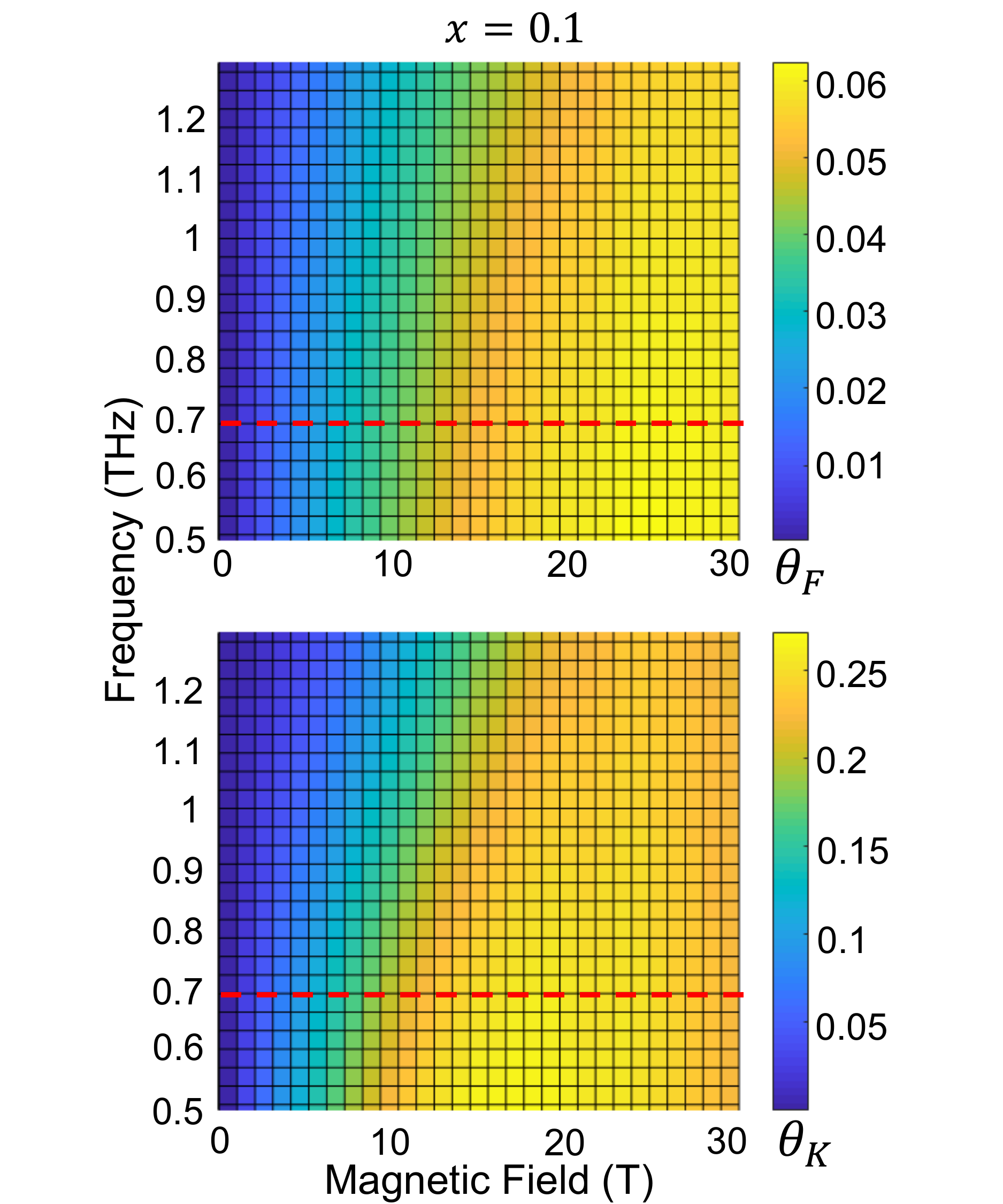}
		\caption{\small Calculated magneto-optical response of $\text{Bi}_{0.9}\text{Sb}_{0.1}$ at $T$ = 21~K in $B$ up to 30~T. (a)~Faraday rotation and (b)~Kerr rotation maps versus THz frequency and magnetic field. The cuts marked by the red dashed lines are plotted together with experimental data in Figs.\,\ref{TIrotation30T}(c) and (d).}
		\label{TIcalmap}
	\end{center}
\end{figure}

Finally, the procedure for calculating the conductivity tensor for the $\bar{\text{M}}$-point electron pocket, $\sigma^{(e,\bar{\text{M}})}_{xx}$ and $\sigma^{(e,\bar{\text{M}})}_{xy}$, is similar to that of the hole pocket. 

We summed up the contributions from all pockets to obtain the total conductivity tensor elements, $\sigma^\text{tot}_{xx}$ and $\sigma^\text{tot}_{xy}$, and calculated the Faraday and Kerr rotations following the process discussed in Section~\ref{polarimetry}. All band parameters are either given in the literature or can be obtained through fits to predetermined band structures. The three free parameters we can tune to match the theoretical results with experimental data are the $\bar{\Gamma}$-point electron scattering rate $\gamma^{\bar{\Gamma}}_e$, the hole scattering rate $\gamma_h$, and the $\bar{\text{M}}$-point electron scattering rate $\gamma^{\bar{\text{M}}}_e$. We found that $\gamma^{\bar{\Gamma}}_e=10\ \text{meV}$, $\gamma_h=60\ \text{meV}$, and $\gamma^{\bar{\text{M}}}_e=10.5\ \text{meV}$ give the best fit with the experimental data, as shown by the polarization rotation curves at 0.7~THz in Figs.\,\ref{TIrotation30T}(c) and (d). Maps of calculated Faraday and Kerr rotations as a function of THz frequency and magnetic field using the optimized parameters are shown in Fig.\,\ref{TIcalmap}. These results can be compared to the experimental maps shown in Figs.\,\ref{TIrotation30T}(a) and (b), and there is agreement between the experimental and theoretical results. 

The three optimized scattering rates, $\gamma^{\bar{\Gamma}}_e$, $\gamma_h$, and $\gamma^{\bar{\text{M}}}_e$, are all larger than the bandwidth of our THz probe, suggesting that the surface carriers do not have high enough mobility for their cyclotron resonance peaks to appear in our measurements. In addition, $\gamma_h$ is much larger than $\gamma^{\bar{\Gamma}}_e$ and $\gamma^{\bar{\text{M}}}_e$. This observation is not surprising as we examine the linewidths of surface bands measured by ARPES in previous studies~\cite{BeniaetAl15PRB}, but its effect is that the contribution of surface holes in the THz polarization rotation signal is negligibly small compared to that of the surface electrons. 

\begin{figure}[htb]
	\begin{center}
		\includegraphics[scale=0.55]{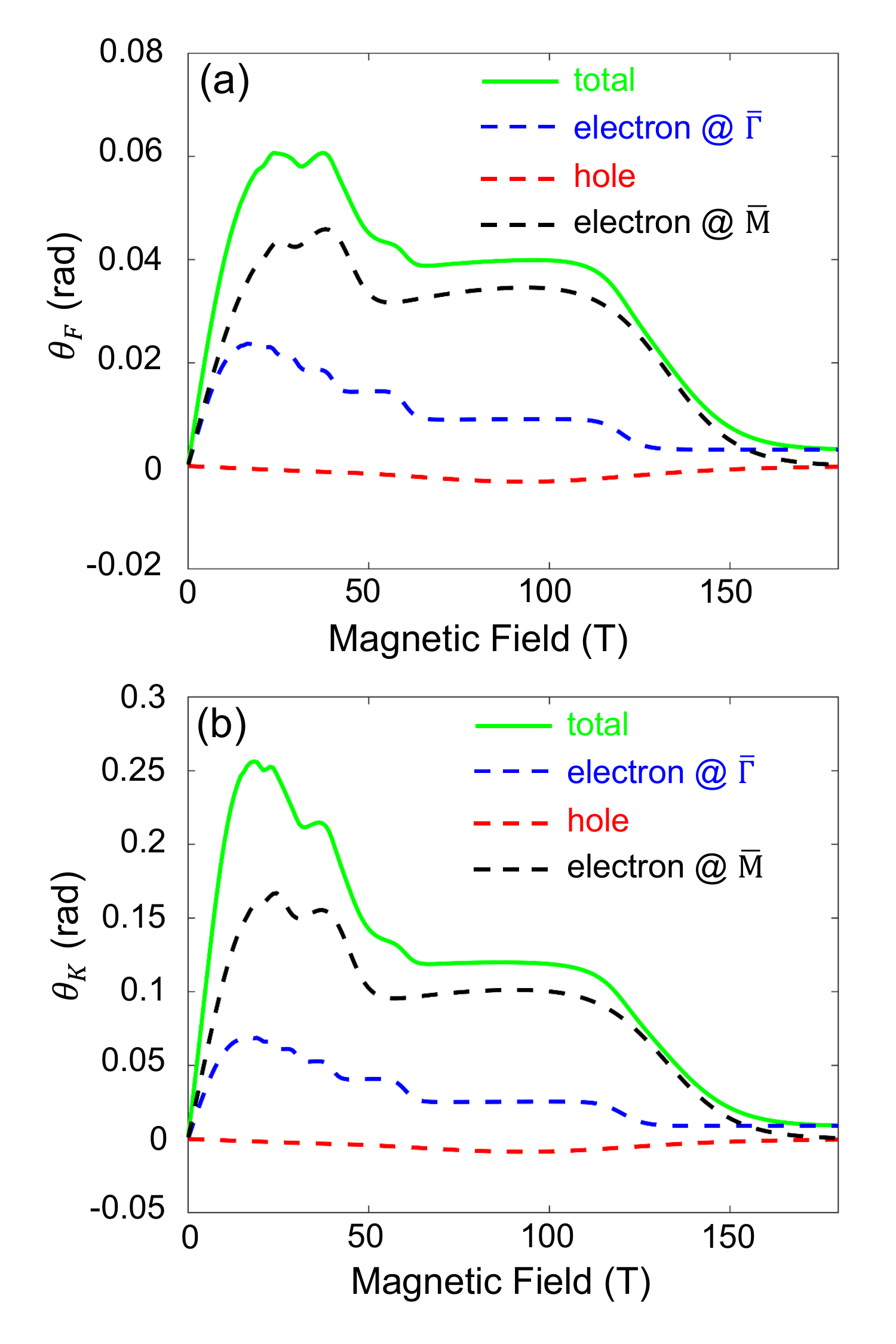}
		\caption{\small Calculated magneto-optical response of the nominal $\text{Bi}_{0.9}\text{Sb}_{0.1}$ film up to 180~T. (a)~Faraday rotation and (b)~Kerr rotation versus magnetic field at a fixed THz frequency of 0.7~THz. The total rotation signal is plotted together with the separate contributions from the three carrier pockets.}
		\label{TIultrahighB}
	\end{center}
\end{figure}

We provide some additional comments on the saturation behavior of Faraday and Kerr rotations in the $B>15\ \text{T}$ region in Figs.\,\ref{TIrotation30T}(c) and (d). It might be tempting to explain these features as the quantized optical Hall effect observed in several recent studies on other TI systems~\cite{WuetAl16Science,OkadaetAl16NC,DziometAl17NC}. However, as shown in the calculation of $\theta_\mathrm{F}$ and $\theta_\mathrm{K}$ in a much wider magnetic field range in Fig.\,\ref{TIultrahighB}, we found that the saturation behavior observed within the $30$~T magnetic field range arises from the summation of the broadened cyclotron resonance signals contributed by the two electron pockets. Theoretical calculations predict a major quantum Hall plateau in the $60\ \text{T}<B<110\ \text{T}$ range as the filling factors of the two electron pockets are both small, but experimental observation might still be challenging because the short carrier localization length in optical experiments (compared to DC experiments) significantly shrinks the quantum Hall plateaus~\cite{Aoki87RPP}. When $B>150\ \text{T}$, the filling factors of the carrier pockets with parabolic dispersions tend to zero, but the Dirac electron pocket at the $\bar{\Gamma}$ point gives a finite signal due to the nontrivial Berry's phase created by electrons circling around in momentum space.  

\section{Summary}

In summary, we performed THz Faraday and Kerr rotation spectroscopy measurements on $\text{Bi}_{1-x}\text{Sb}_{x}$ thin films.  This alloy system exhibits a semimetal-to-topological-insulator transition as a function of $x$.  By using single-shot time-domain THz spectroscopy combined with a 30-T table-top pulsed magnet, we observed distinctly different behaviors between semimetallic ($x < 0.07$) and topological insulator ($x > 0.07$) samples.  We were able to distinguish the origins of the magneto-optical responses of these films by comparing experimental data with predictions from our theoretical models.  We found that a surface (bulk) band model including some material parameters established in previous studies can completely explain the THz Hall signal of all samples. The combined effort of the THz polarimetry experiments performed in high magnetic fields and the detailed theoretical analysis can be applied to other topological materials to investigate surface and bulk carrier contributions to the optical conductivity.

\begin{acknowledgments}
This research was primarily supported by the National Science Foundation through the Center for Dynamics and Control of Materials: an NSF MRSEC under Cooperative Agreement No.\ DMR-1720595.  I.K.\ acknowledges support from the Japan Society for the Promotion of Science (JSPS) through the Bilateral Joint Research Project.  I.K.\ and J.T.\ thank the Ministry of Education, Culture, Sports, Science and Technology (MEXT)/JSPS for support through KAKENHI Grant Nos.\ 16H06010, 17H06124, and 18H04288. G.A.F.\ gratefully acknowledges support from a Simons Fellowship.
\end{acknowledgments}

% Create the reference section using BibTeX:
%\clearpage
%\bibliography{jun}

%apsrev4-2.bst 2019-01-14 (MD) hand-edited version of apsrev4-1.bst
%Control: key (0)
%Control: author (72) initials jnrlst
%Control: editor formatted (1) identically to author
%Control: production of article title (-1) disabled
%Control: page (0) single
%Control: year (1) truncated
%Control: production of eprint (0) enabled
%

\end{document}